\ifx\arxiv\undefined
\documentclass[times,twocolumn,final]{elsarticle}
\else
\documentclass[times,twocolumn,final,nopreprintline]{elsarticle}
\fi
\ifx\arxiv\undefined
\usepackage{medima}
\else
\RequirePackage{geometry}
\fi
\usepackage{framed,multirow}
\usepackage[utf8]{inputenc} % allow utf-8 input
\usepackage[T1]{fontenc}    % use 8-bit T1 fonts
\usepackage{amsfonts}
\usepackage{nicefrac}
\usepackage{microtype}
\usepackage{stfloats}
\usepackage{csquotes}
\usepackage{xspace}

\usepackage{amssymb}
\usepackage{latexsym}
\usepackage{gensymb}

\usepackage{url}
\usepackage{xcolor}

\usepackage[%
breaklinks=true,
colorlinks=true,
bookmarksnumbered=true% Lesezeichen auch nummerieren
]{hyperref}
\usepackage[labelfont=bf,font=small]{caption}% Figure 1 in bold (https://tex.stackexchange.com/questions/32459/figure-how-to-have-figure-1-5-in-bold), smaller caption font (font size overview: https://tex.stackexchange.com/questions/24599/what-point-pt-font-size-are-large-etc)

\usepackage{booktabs}

\usepackage[official]{eurosym}
\usepackage{pdfpages}
\usepackage{stackengine}
\usepackage{graphicx}
\usepackage{lscape}

\usepackage{booktabs}
\usepackage{float}
\usepackage{framed,multirow}
\usepackage{setspace}
\usepackage{threeparttable}
\usepackage{tabularx}
\usepackage{amsmath}% http://ctan.org/pkg/amsmath
\usepackage{siunitx}
\DeclareSIUnit\px{px}

\newcolumntype{L}[1]{>{\raggedright\arraybackslash}p{#1}} % linksbündig mit Breitenangabe
\newcolumntype{C}[1]{>{\centering\arraybackslash}p{#1}} % zentriert mit Breitenangabe
\newcolumntype{R}[1]{>{\raggedleft\arraybackslash}p{#1}} % rechtsbündig mit Breitenangabe

\usepackage{amssymb}
\usepackage{latexsym}

\usepackage{xcolor}
\usepackage{booktabs}
\usepackage{enumitem}
\usepackage{ifthen}

\usepackage[official]{eurosym}
\usepackage{pdfpages}
\usepackage{stackengine}
\usepackage{algorithm}
\usepackage{algorithmic}
\usepackage[labelformat=simple]{subcaption}
\makeatletter
\renewcommand\p@subfigure{\thefigure\,}
\makeatother

\usepackage{xcolor}
\definecolor{expert-col}{RGB}{18, 102, 35}
\usepackage{mathrsfs}
\usepackage{pifont}
\usepackage{mhchem}

\usepackage{acro}
\acsetup{make-links=true}
\newcommand{\acrocolor}[1]{\textcolor{black}{#1}}% Links but no colors for acronyms
\DeclareAcronym{hsi}{short=\acrocolor{HSI},long=hyperspectral imaging}
\DeclareAcronym{msi}{short=\acrocolor{MSI},long=multispectral imaging}
\DeclareAcronym{tpi}{short=\acrocolor{TPI},long=tissue parameter images}
\DeclareAcronym{dsc}{short=\acrocolor{DSC},long=dice similarity coefficient}
\DeclareAcronym{asd}{short=\acrocolor{ASD},long=average surface distance}
\DeclareAcronym{nsd}{short=\acrocolor{NSD},long=normalized surface dice}
\DeclareAcronym{sto2}{short=\acrocolor{\ce{StO2}},long=tissue oxygen saturation}
\DeclareAcronym{npi}{short=\acrocolor{NPI},long=near-infrared perfusion index}
\DeclareAcronym{twi}{short=\acrocolor{TWI},long=tissue water index}
\DeclareAcronym{thi}{short=\acrocolor{THI},long=tissue hemoglobin index}
\DeclareAcronym{elu}{short=\acrocolor{ELU},long=exponential linear unit}
\DeclareAcronym{gpu}{short=\acrocolor{GPU},long=graphics processing unit}
\DeclareAcronym{sd}{short=\acrocolor{SD},long=standard deviation}
\DeclareAcronym{ce}{short=\acrocolor{CE},long=cross-entropy}
\DeclareAcronym{json}{short=\acrocolor{JSON}, long=JavaScript object notation}
\DeclareAcronym{tsne}{short=\acrocolor{$t$-SNE}, long=$t$-distributed stochastic neighbour approach}
\DeclareAcronym{svm}{short=\acrocolor{SVM}, long=support vector machine}

\definecolor{dkfzgreen}{rgb}{0,0.6,0.2}
\definecolor{dkfzyellow}{rgb}{1,0.839,0}
\definecolor{dkfzorange}{rgb}{1,0.361,0}
\definecolor{mialinkcolor}{RGB}{0,128,172}

\newcommand{\oname}[1]{\textit{#1}}
\newcommand{\autocite}[1]{\citep{#1}}
\newcommand{\varAcquisitionTime}{seven seconds\xspace}
\newcommand{\varGeneralizationPhrasing}{new surgery\xspace}
\newcommand{\varGeneralizationPhrasingS}{new surgeries\xspace}

\newcommand{\varPixelHSITotalWeights}{\num{34275}\xspace}
\newcommand{\varPixelTPITotalWeights}{\num{27819}\xspace}
\newcommand{\varPixelRGBTotalWeights}{\num{27619}\xspace}
\newcommand{\varTotalImages}{506\xspace}
\newcommand{\varTotalPigs}{20\xspace}
\newcommand{\varTotalTrainingPigs}{15\xspace}
\newcommand{\varTotalTrainingImages}{340\xspace}
\newcommand{\varMaxTrainingPigsDatasetSize}{14\xspace}
\newcommand{\varTotalClassesDatasetSize}{eight\xspace}
\newcommand{\varClassesDatasetSize}{\oname{background}, \oname{stomach}, \oname{small intestine}, \oname{colon}, \oname{liver}, \oname{spleen}, \oname{skin} and \oname{peritoneum}\xspace}
\newcommand{\varTotalOrganClasses}{18\xspace}
\newcommand{\varTotalClasses}{19\xspace}
\newcommand{\varMinPigsOrgan}{5\xspace}
\newcommand{\varMaxPigsOrgan}{20\xspace}
\newcommand{\varMinImagesOrgan}{32\xspace}
\newcommand{\varMaxImagesOrgan}{405\xspace}
\newcommand{\varInvalidImages}{221\xspace}
\newcommand{\varBackgroundPixelRatio}{\SI{47}{\percent} (\ac{sd} \SI{ 24}{\percent})\xspace}
\newcommand{\varInvalidPixelRatio}{\SI{2}{\percent} (\ac{sd} \SI{ 3}{\percent})\xspace}

\newcommand{\varEpochTable}{\begin{tabular}{llll}
\toprule
model & \# pixels & epoch size & batch size \\
\midrule
image & \num{307200} & \num{500} & \num{5} \\
patch\_64 & \num{4096} & \num{37632} & \num{336} \\
patch\_32 & \num{1024} & \num{150528} & \num{1176} \\
superpixel & $\approx 300$ & \num{500760} & \num{1560} \\
pixel & \num{1} & \num{153608400} & \num{118800} \\
\bottomrule
\end{tabular}\xspace}
\newcommand{\varPixelImprovement}{\SI{80.4}{\percent}\xspace}
\newcommand{\varSuperpixelImprovement}{\SI{11.9}{\percent}\xspace}
\newcommand{\varPatchThreeTwoImprovement}{\SI{8.1}{\percent}\xspace}
\newcommand{\varPatchSixFourImprovement}{\SI{5.5}{\percent}\xspace}
\newcommand{\varImageImprovement}{\SI{2.8}{\percent}\xspace}
\newcommand{\varRaterInterAdditional}{8\xspace}
\newcommand{\varRaterInterMissing}{7\xspace}
\newcommand{\varRaterInterTotalMaskDiffPixels}{\SI{34063}{\px}\xspace}
\newcommand{\varRaterInterTotalMaskDiffImages}{14\xspace}
\newcommand{\varRaterInterDSC}{0.89 (\ac{sd} 0.07)\xspace}
\newcommand{\varRaterInterASD}{4.88 (\ac{sd} 5.33)\xspace}
\newcommand{\varRaterInterNSD}{0.80 (\ac{sd} 0.08)\xspace}
\newcommand{\varRaterIntraAdditional}{6\xspace}
\newcommand{\varRaterIntraMissing}{4\xspace}
\newcommand{\varRaterIntraTotalMaskDiffPixels}{\SI{37397}{\px}\xspace}
\newcommand{\varRaterIntraTotalMaskDiffImages}{14\xspace}
\newcommand{\varRaterIntraDSC}{0.91 (\ac{sd} 0.05)\xspace}
\newcommand{\varRaterIntraASD}{4.74 (\ac{sd} 5.04)\xspace}
\newcommand{\varRaterIntraNSD}{0.82 (\ac{sd} 0.06)\xspace}
\newcommand{\varRaterTotalImages}{20\xspace}
\newcommand{\varSeedVariationTestDSC}{[0.893; 0.900]\xspace}
\newcommand{\varSeedVariationTestASD}{[5.542; 6.948]\xspace}
\newcommand{\varSeedVariationTestNSD}{[0.796; 0.806]\xspace}
\newcommand{\varTrainingTimeTotal}{\SI{292}{\hour}\xspace}
\newcommand{\varSpxLimitDSC}{0.92 (\ac{sd} 0.03)\xspace}
\newcommand{\varSpxLimitASD}{2.91 (\ac{sd} 0.74)\xspace}
\newcommand{\varSpxLimitNSD}{0.74 (\ac{sd} 0.04)\xspace}

\newcommand{\varPerformanceSuperpixelDSC}{0.82 (\ac{sd} 0.06)\xspace}
\newcommand{\varPerformanceSuperpixelASD}{16.51 (\ac{sd} 9.23)\xspace}
\newcommand{\varPerformanceSuperpixelNSD}{0.61 (\ac{sd} 0.09)\xspace}

\newcommand{\varPerformanceImageDSC}{0.90 (\ac{sd} 0.04)\xspace}
\newcommand{\varPerformanceImageASD}{6.19 (\ac{sd} 3.20)\xspace}
\newcommand{\varPerformanceImageNSD}{0.80 (\ac{sd} 0.07)\xspace}
\newcommand{\varCMThreshold}{\SI{95}{\percent}\xspace}
\newcommand{\varCMTotalClassesAboveThreshold}{8\xspace}
\newcommand{\varCMVenaCavaSensitivity}{\SI{57.1}{\percent}\xspace}
\newcommand{\varCMVenaCavaMaxConfusion}{\SI{32.2}{\percent}\xspace}

\newcommand{\varCMVenaCavaTotalImages}{32\xspace}
\newcommand{\varCMVenaCavaPixels}{\SI{4192}{\px} (\ac{sd} \SI{3621}{\px})\xspace}

\newcommand{\varThresholdsAggregation}{mean\xspace}
\newcommand{\varThresholdsAggregationLow}{\SI{20}{\px}\xspace}
\newcommand{\varThresholdsAggregationHigh}{\SI{80}{\px}\xspace}
\newcommand{\varThresholdsAggregationSkinFactor}{2.5\xspace}
\newcommand{\varLrDSCDiff}{0.007\xspace}
\newcommand{\varLrDefault}{0.001\xspace}
\newcommand{\varLrHigher}{0.01\xspace}
\newcommand{\varLrLower}{0.0001\xspace}
\newcommand{\varInferenceTime}{\SI{122}{\ms}\xspace}
\definecolor{Pixel}{HTML}{FF8C00}
\definecolor{Superpixel}{HTML}{E6003D}
\definecolor{PatchThreeTwo}{HTML}{4169E1}
\definecolor{PatchSixFour}{HTML}{808000}
\definecolor{Image}{HTML}{800080}

\ifx\arxiv\undefined
\journal{Medical Image Analysis}
\verso{Seidlitz \& Sellner \textit{et~al.}}
\else
\def\KWD{}
\newcommand{\snm}[1]{#1}
\usepackage{fancyhdr}
\fi

\begin{document}
% Must be set here to have an effect
\hypersetup{allcolors=mialinkcolor}

\begin{frontmatter}

\title{Robust deep learning-based semantic organ segmentation in hyperspectral images}%

\author[1,2]{Silvia \snm{Seidlitz}}\corref{cor1}
\ead{s.seidlitz@dkfz-heidelberg.de}
\author[1,2]{Jan \snm{Sellner}\corref{cor1}}
\ead{j.sellner@dkfz-heidelberg.de}
\author[1,3]{Jan \snm{Odenthal}}
\author[3,4]{Berkin \snm{\"Ozdemir}}
\author[3,4]{Alexander \snm{Studier-Fischer}}
\author[3,4]{Samuel \snm{Kn\"odler}}
\author[1,4]{Leonardo \snm{Ayala}}
\author[1,6]{Tim J. \snm{Adler}}
\author[2,3]{Hannes G. \snm{Kenngott}}
\author[1]{Minu \snm{Tizabi}}
\author[2,3,4]{Martin \snm{Wagner}}
\author[2,3,4]{Felix \snm{Nickel}}
\author[3,4]{Beat P. \snm{M\"uller-Stich}}
\author[1,2,4,5,6]{Lena \snm{Maier-Hein}}
\cortext[cor1]{First authors (*) contributed equally to this paper and the order was assigned randomly.}

\address[1]{Division of Intelligent Medical Systems, German Cancer Research Center (DKFZ), Heidelberg, Germany}
\address[2]{Helmholtz Information and Data Science School for Health, Karlsruhe/Heidelberg, Germany}
\address[3]{Department of General, Visceral, and Transplantation Surgery, Heidelberg University Hospital, Heidelberg, Germany}
\address[4]{Medical Faculty, Heidelberg University, Heidelberg, Germany}
\address[5]{HIP Helmholtz Imaging Platform, German Cancer Research Center (DKFZ), Heidelberg, Germany}
\address[6]{Faculty of Mathematics and Computer Science, Heidelberg University, Heidelberg, Germany}

\ifx\arxiv\undefined
\received{14 November 2021}
\finalform{28 March 2022}
\accepted{20 May 2022}
\availableonline{27 May 2022}
\fi


\begin{abstract}
%Purpose
Semantic image segmentation is an important prerequisite for context-awareness and autonomous robotics in surgery. The state of the art has focused on conventional RGB video data acquired during minimally invasive surgery, but full-scene semantic segmentation based on spectral imaging data and obtained during open surgery has received almost no attention to date. 
%Methods
To address this gap in the literature, we are investigating the following research questions based on \ac{hsi} data of pigs acquired in an open surgery setting: (1) What is an adequate representation of \ac{hsi} data for neural network-based fully automated organ segmentation, especially with respect to the spatial granularity of the data (pixels \textit{vs.} superpixels \textit{vs.} patches \textit{vs.} full images)? (2) Is there a benefit of using \ac{hsi} data compared to other modalities, namely RGB data and processed \ac{hsi} data (e.g. tissue parameters like oxygenation), when performing semantic organ segmentation? 
%Results
According to a comprehensive validation study based on \varTotalImages \ac{hsi} images from \varTotalPigs pigs, annotated with a total of \varTotalClasses classes, deep learning-based segmentation performance increases --- consistently across modalities --- with the spatial context of the input data. Unprocessed \ac{hsi} data offers an advantage over RGB data or processed data from the camera provider, with the advantage increasing with decreasing size of the input to the neural network. Maximum performance (\ac{hsi} applied to whole images) yielded a mean \ac{dsc} of \varPerformanceImageDSC, which is in the range of the inter-rater variability (\ac{dsc} of \varRaterInterDSC).
%Conclusion
We conclude that \ac{hsi} could become a powerful image modality for fully-automatic surgical scene understanding with many advantages over traditional imaging, including the ability to recover additional functional tissue information.
Our code and pre-trained models are available at \href{https://github.com/IMSY-DKFZ/htc}{https://github.com/IMSY-DKFZ/htc}.
\end{abstract}

\end{frontmatter}

\acresetall

\begin{figure*}[htb]
    \centering
    \includegraphics[width=\textwidth]{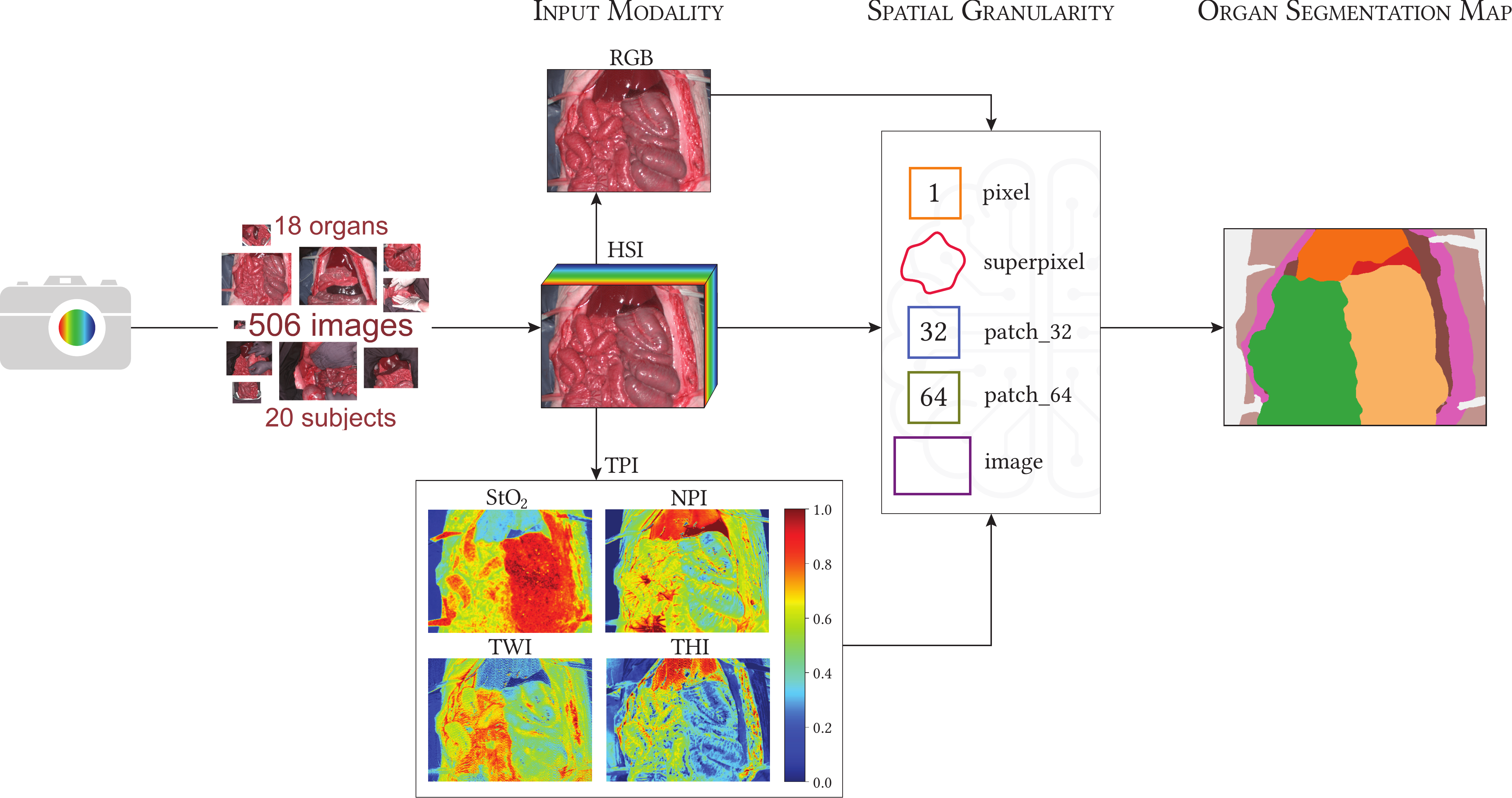}
    \caption[Pipeline overview]{In this paper, we address the challenge of semantic scene segmentation based on \acf*{hsi} data. We put a particular focus on the optimal representation of the data, both in the spectral domain (here referred to as \textit{input modality}) and in the spatial domain (here referred to as \textit{spatial granularity}). Modalities comprise \acf*{hsi} ($c=100$ channels) as well as corresponding RGB data ($c=3$) and \acf*{tpi} ($c=4$). The latter includes \acf*{sto2}, \acf*{npi}, \acf*{twi} and \acf*{thi}. Each combination (input modality, spatial granularity) yields an organ segmentation map for the whole image after potential aggregation of image parts. The corresponding neural network architectures are illustrated in \autoref{fig:model_overview}.}
    \label{fig:intro}
\end{figure*}

\section{Introduction}

Surgical data science is the discipline of capturing, organizing, analysing and modelling surgical data in order to improve the quality of interventional healthcare \autocite{maier-hein_surgical_2017, maier-hein_surgical_2021}. Semantic scene segmentation is an important prerequisite for various tasks in surgical data science, including context-aware assistance and surgical robotics. So far, the scientific literature has focused on binary segmentation tasks (e.g. medical instruments \autocite{ros_comparative_2021}) and conventional RGB video data (see e.g. \autocite{scheikl_deep_2020, grammatikopoulou_cadis_2021}).

Recently, spectral imaging \autocite{clancy_surgical_2020} has evolved as a promising technique for advanced optical imaging in the operating room. While conventional RGB imaging is limited by imitating the human eye, spectral imaging systems capture the reflectance spectrum of the tissue over the entire field of view, thereby generating a datacube consisting of two spatial and one spectral dimension. Since the underlying tissue optical properties determine the measured tissue reflectance spectrum, spectral imaging has the potential to extract biological information such as tissue type and pathologies while being non-invasive \autocite{studier-fischer_spectral_2021}.
Spectral imaging with up to tens of spectral bands is generally referred to as \ac{msi}, whereas spectral imaging with up to hundreds of spectral bands is named \ac{hsi} \autocite{clancy_surgical_2020}. Early work on semantic scene segmentation with \ac{msi} data indicated that more detailed spectral information may improve the segmentation performance, but the number of classes was relatively low with six and three classes, respectively \autocite{moccia_uncertainty-aware_2018, garifullin_hyperspectral_2018}. It has not been determined yet whether there is a benefit in using \ac{hsi} data over processed \ac{hsi} data (e.g. in the form of parameter images, cf. \autoref{fig:intro}) and RGB data. Furthermore, we are not aware of any prior work that investigated what is an adequate representation of medical \ac{hsi} data for neural network-based fully automated scene segmentation, especially with respect to the spatial granularity of the data. The spatial variability of images acquired during open surgery is large due to the less controlled imaging conditions (e.g. imaging distance), complex three-dimensional relationships between multiple soft tissues as well as anatomical differences between individuals. In addition, acquisition and annotation of large data sets are difficult and time-consuming. In order to determine the ideal spatial granularity of the \ac{hsi} input data, the generalization capability towards a \varGeneralizationPhrasing (e.g. unseen individual or experimental conditions) and the required amount of training data thus also need to be considered.

To address this gap in the literature, this paper investigates the following research questions (cf. \autoref{fig:intro} for an overview):
\begin{enumerate}[noitemsep]
    \item What is an adequate representation of \ac{hsi} data for neural network-based fully automated organ segmentation? Specifically, what is the optimal granularity of the data (pixels \textit{vs.} superpixels \textit{vs.} patches \textit{vs.} full images) with respect to segmentation quality, the required number of training cases and the capability to generalize towards \varGeneralizationPhrasingS? 
    \item Is there a benefit of using \ac{hsi} data compared to RGB data and processed \ac{hsi} data (e.g. tissue parameter estimations) when performing semantic organ segmentation? 
\end{enumerate}

The remainder of this paper is structured as follows: After presenting the related work in \autoref{sec:related_work}, we describe our \ac{hsi} data set and approach to semantic organ segmentation in \autoref{sec:methods}. The performed experiments and results are presented in \autoref{sec:experiments} and summarized and discussed in \autoref{sec:discussion}.

\section{Related work}\label{sec:related_work}

As only very limited prior work on automatic organ segmentation in \ac{msi}/\ac{hsi} exists, we first summarize related work on organ segmentation based on RGB data in surgery and then present a brief overview on the state of the art in segmentation with \ac{msi}/\ac{hsi} data, both within and outside the field of biomedical image analysis.

\subsection{Deep learning-based organ segmentation on RGB data}

During the past years, deep learning-based segmentation of RGB data has found several applications in surgery, especially in minimally invasive surgery \autocite{rivas-blanco_review_2021} such as cataract \autocite{grammatikopoulou_cadis_2021} or colorectal surgery \autocite{maier-hein_heidelberg_2021}. However, work has mainly focused on medical instrument segmentation, driven by various challenges in this area (e.g. CATARACTS challenge on automatic tool segmentation in cataract surgery \autocite{al_hajj_cataracts_2019}, Robust Medical Instrument Segmentation challenge in laparoscopic surgery \autocite{maier-hein_heidelberg_2021}). Only few recent works have tackled organ segmentation, either restricted to organ classes (e.g. \autocite{fu_more_2019, webster_deep_2017}) or, more often, in the context of full scene segmentation (e.g. \autocite{allan_2018_2020, zhou_feature_2019, madad_zadeh_surgai_2020, scheikl_deep_2020}). The data sets used differ highly in terms of annotation sparsity (e.g. full scene \textit{vs.} specific organ segmentation) and the number of considered classes. Input to the models are video frames of varying size (e.g. $960 \times 540$ in \autocite{scheikl_deep_2020}, $512 \times 512$ in \autocite{laves_dataset_2019}). Relatively few works have tackled organ segmentation in open surgery, where, compared to minimally invasive surgery, image acquisition is often more difficult to realize and challenges arise from the even larger complexity and variability of the surgical scene \autocite{gong_using_2021}. We are aware of only a single investigation of deep learning-based organ segmentation on RGB images in open surgery: \citeauthor{gong_using_2021} analysed segmentation performance under different imaging conditions such as lightning changes or varying distances based on RGB images of 130 patients and found that these factors have a high influence on the image scores. 

Overall, prior work has identified several major challenges for automated organ segmentation on RGB data such as a high variability in tissue appearance across patients (e.g. \autocite{webster_deep_2017, collins_segmenting_2015}) and across images (e.g. due to occlusions or deformations \autocite{moccia_uncertainty-aware_2018}) as well as the variability in the image acquisition. Including further spectral information could be key for addressing those challenges as \ac{msi}/\ac{hsi} may be less reliant on the spatial context and encodes additional clinical information such as tissue perfusion \autocite{fei_chapter_2020}.

\subsection{Segmentation with \ac{msi}/\ac{hsi} data}

\paragraph{Within biomedical image analysis} Only a small number of papers address a biomedical segmentation problem based on \ac{msi}/\ac{hsi} data with deep learning \autocite{khan_trends_2021}. Even without restricting the search to deep learning-based approaches, we could only identify nine related publications (with only four of them using deep learning techniques):

\citeauthor{trajanovski_tongue_2021} segmented healthy and tumorous tongue tissue in histopathological \ac{hsi} images on an in-house data set consisting of 14 patients (one image per patient) \autocite{trajanovski_tongue_2021}. Expanding on their earlier work \autocite{trajanovski_tumor_2019}, they compared several pixel-based networks, networks based on patches of size $256 \times 256$ and hybrid networks, taking a combination of entire pixel spectra and patches with a reduced number of channels as input. They found a U-Net architecture \autocite{ronneberger_u-net_2015} based on patches to perform best in their specific segmentation task. However, as the performance analysis was conducted on the validation data set (on which hyperparameters were tuned), a subsequent evaluation on an independent test set remains to be performed.

\citeauthor{garifullin_hyperspectral_2018} analysed 55 retinal \ac{msi} images and segmented three tissue types (vessels, optic disc and macula) \autocite{garifullin_hyperspectral_2018}. They used SegNet \autocite{badrinarayanan_segnet_2016} and Dense-FCN \autocite{jegou_one_2017} models and compared \ac{msi} with RGB data but their results did not reveal a clear winner (neither from the model nor the modality perspective).

\citeauthor{cervantes-sanchez_automatic_2021} analysed 18 \ac{hsi} images from seven hepatic surgery patients and 21 \ac{hsi} images from seven thyroid surgery patients. They created sparse annotations of circular shape for four organs (liver, bile duct, artery, portal vein) in the case of hepatic surgeries and three organs (thyroid, parathyroid, muscle) in the case of thyroid surgeries \autocite{cervantes-sanchez_automatic_2021}. They compared the performance of several machine learning methods (logistic regression \autocite{cox_regression_1958}, \ac{svm} \autocite{boser_training_1992}, multilayer perceptron \autocite{haykin_neural_1994} and U-Net) based on single pixels or small patches of an $8 \times 8$ shape for automatic segmentation of the annotated organ classes. However, evaluation was only performed on the sparse annotations and on the validation data set (on which hyperparameters were tuned). Therefore, a subsequent evaluation on full semantic annotations and an independent test set remains to be performed.

In the project HELICoiD \autocite{fabelo_helicoid_2016}, the potential of \ac{hsi} for the segmentation of tumorous and healthy brain tissue from patients undergoing neurosurgery was studied. The entire data set comprised 36 images from 22 patients and was made publicly available \autocite{fabelo_-vivo_2019}. Sparse annotations of four classes (tumor tissue, normal brain tissue, blood vessels and background) were created by combining manual expert segmentations based on pathological findings with $k$-means clustering ($k=15$). Pixels belonging to small clusters were removed as they were suspected to be annotation errors. \citeauthor{ravi_manifold_2017} trained a Semantic Texton Forest \autocite{shotton_semantic_2008} on a subset of the HELICoiD data set, consisting of 33 \ac{hsi} brain images from 18 patients that were embedded with an adapted version of the \ac{tsne} \autocite{maaten_visualizing_2008} to segment tumorous and healthy brain tissue \autocite{ravi_manifold_2017}. However, the performance analysis was conducted on the validation data set (on which hyperparameters were tuned). \citeauthor{fabelo_spatio-spectral_2018} proposed a multi-class semantic segmentation concept based on fusing a segmentation prediction from a supervised pixel-based \ac{svm} classifier that was spatially homogenized through $k$-nearest neighbours filtering with a segmentation prediction obtained through unsupervised clustering \autocite{fabelo_spatio-spectral_2018}. Due to the sparsity of the annotations, a quantitative validation of the segmentations could only be performed for the \ac{svm} classifier but the separation between train, validation and test data is unclear. \citeauthor{fabelo_surgical_2019} used 26 \ac{hsi} brain images from 16 patients (6 patients with grade IV glioblastoma and 10 patients with normal brain tissue) to compare baseline \ac{svm}-based methods to a pixel-based deep neural network and a two dimensional convolutional neural network classifier on small patches of an $11 \times 11$ shape \autocite{fabelo_surgical_2019}. They found that both deep learning-based methods yielded similar performance and outperformed the \ac{svm}-based methods. However, the performance analysis was conducted on the validation data set (on which hyperparameter tuning was performed). Given the sparsity of the annotations and unclear or missing separation between train, validation and test data, a subsequent performance assessment on an independent test set with full semantic annotations remains to be performed for all three studies.

\citeauthor{moccia_uncertainty-aware_2018} acquired \ac{msi} data of seven pigs (57 images) in the setting of hepatic laparoscopic surgery \autocite{moccia_uncertainty-aware_2018}. They turned the actual organ segmentation problem into a classification problem in the following way: Based on manually extracted textural and spectral features from automatically segmented superpixels, they trained a \ac{svm} to classify six organs (liver, gallbladder, spleen, diaphragm, intestine, and abdominal wall). They showed that the classification accuracy for their \ac{msi} data was superior to the classification accuracy for a selection of only three channels. However, the selected channels were too narrow to represent an RGB image.

\citeauthor{akbari_wavelet-based_2008} acquired seven \ac{hsi} images of abdominal organs for a single pig. Five organs (spleen, colon, small intestine, bladder, peritoneum) were annotated for these images and pixel-based organ classification was performed while learning vector quantization \autocite{kohonen_learning_1995} of compressed spectra \autocite{akbari_wavelet-based_2008}. Given the small size of the data set consisting only of a single individual and the unclear separation between train and test data, a subsequent evaluation on an independent test set comprising a larger number of individuals remains to be performed.

\paragraph{Outside biomedical image analysis} \ac{msi}/\ac{hsi} is applied in various fields such as biochemistry, agriculture, archaeology and especially remote sensing \autocite{prasad_hyperspectral_2020}. However, investigation of deep learning-based semantic scene segmentation in those fields is rare \autocite{khan_modern_2018}. The validity of existing works is very limited due to small data sets composed of only one to two images and training and testing being performed on the same data (e.g. \autocite{alam_crf_2016, mughees_efficient_2016, nalepa_towards_2020, paul_classification_2021}). Generally, the application of deep learning-based semantic scene segmentation is hampered by limitations in the available annotations \autocite{vali_deep_2020}: training data is sparse and often only several discrete pixels instead of entire images are labelled, while due to the high dimensionality of the data, large data sets would be required to avoid overfitting \autocite{zhu_spectral-spatial-dependent_2021}. Due to these limitations, most segmentation tasks in these fields are addressed via pixel-based classification (e.g. \autocite{nalepa_towards_2020}) and the few existing patch-based or image-based segmentation approaches are at high risk of train-test-leakage \autocite{nalepa_validating_2019}. \\

In summary, prior work on semantic scene segmentation in open surgery is extremely sparse and even non-existent in medical \ac{hsi}. Furthermore, the data sets used so far are rather small, and the high complexity and variability in surgical scenes due to non-standardized image acquisition, inter-subject variability and complex three-dimensional relationships between multiple soft tissues (e.g. overlapping tissue, shadowing, deformations) \autocite{gong_using_2021} remain to be addressed. Among the related work on organ segmentation from \ac{msi}/\ac{hsi} data, models have been based on superpixels \autocite{moccia_uncertainty-aware_2018}, patches \autocite{trajanovski_tongue_2021, garifullin_hyperspectral_2018, fabelo_surgical_2019, cervantes-sanchez_automatic_2021} and pixels \autocite{akbari_wavelet-based_2008, fabelo_surgical_2019}. However, the optimal granularity of the data with respect to segmentation quality, the required number of training cases, and the capability to generalize towards \varGeneralizationPhrasingS given the large variability in the surgical scene, has not been determined up to the present date. Furthermore, no prior work could show a clear benefit of \ac{msi}/\ac{hsi} data over RGB data for deep learning-based organ segmentation. We address these gaps in the literature based on a semantically annotated \ac{hsi} data set of unprecedented size and number of classes (\varTotalImages images from \varTotalPigs pigs semantically annotated with \varTotalClasses classes): The data sets in the related work are composed of a maximum of 22 individuals \autocite{fabelo_-vivo_2019} and annotated with a maximum of six classes \autocite{moccia_uncertainty-aware_2018}. To the best of our knowledge, the largest medical \ac{hsi} data sets outside the field of segmentation are composed of 316 images from 30 patients, of which 215 images were annotated with 35 classes \autocite{hyttinen_oral_2020} and 9059 images from 46 pigs annotated with 20 classes \autocite{studier-fischer_spectral_2021}. Despite the data sets provided by \autocite{fabelo_-vivo_2019, hyttinen_oral_2020, studier-fischer_spectral_2021} surpassing our data set in the number of individuals, none of these larger data sets provides semantic segmentations.

\section{Materials and methods}
\label{sec:methods}

The following sections describe the hardware and data set that served as a foundation for this work (\autoref{sec:data}) and the individual components of our image processing pipeline (\autoref{sec:image_proc_pipeline}). Our implementation and the pre-trained models can be found in our GitHub repository\footnote{\href{https://github.com/IMSY-DKFZ/htc}{https://github.com/IMSY-DKFZ/htc}} \autocite{sellner_hyperspectral_2022}.

\subsection{Image acquisition and data set}
\label{sec:data}

The \ac{hsi} data was acquired at the Heidelberg University Hospital after approval by the Committee on Animal Experimentation of the regional council Baden-W\"urttemberg in Karlsruhe, Germany (G-161/18 and G-262/19). \ac{hsi} images were taken for \varTotalPigs pigs that were managed according to the German laws for animal use and care and in agreement with the directives of the European Community Council (2010/63/EU). Details on the animals, the performed anaesthesia and surgical procedures are available in \autocite{studier-fischer_spectral_2021}.

\subsubsection{HSI camera system}
\label{sec:hsi_camera}

The \ac{hsi} camera system Tivita\textsuperscript{\textregistered} Tissue (Diaspective Vision GmbH, Am Salzhaff, Germany) was used to acquire the \ac{hsi} data. In a push-broom fashion, it captures hyperspectral images with a spectral resolution of approximately \SI{5}{nm} in the spectral range between \SI{500}{nm} and \SI{1000}{nm}, resulting in datacubes with a size of $640\times480\times100$ (width $w$ $\times$ height $h$ $\times$ number of spectral channels $c$). The camera system images an area of approximately 20 $\times$ \SI{30}{cm}. An imaging distance of about \SI{50}{cm} is ensured through an integrated distance calibration system composed of two light marks that overlap if the distance is correct. The image acquisition takes approximately \varAcquisitionTime. In addition to the \ac{hsi} datacubes, the camera system estimates \ac{tpi} that include oxygenation (\ac{sto2}), perfusion (\ac{npi}), water content (\ac{twi}) and hemoglobin content (\ac{thi}) from the \ac{hsi} datacubes. Furthermore, RGB images are reconstructed from the \ac{hsi} data by aggregating spectral channels capturing red, green and blue light, respectively. The underlying calculations are described in \autocite{holmer_hyperspectral_2018}. \autoref{fig:intro} shows the reconstructed RGB image and \ac{tpi} associated with an exemplary \ac{hsi} datacube. More technical details on the hardware are available in \autocite{holmer_hyperspectral_2018, kulcke_compact_2018}.

\subsubsection{Data acquisition}

In order to prevent distortion of the spectra from stray light, light sources other than the integrated halogen lighting unit were shut off during image acquisition and window blinds were closed. Motion artefacts were reduced in the following ways: (1) The camera was mounted on a swivel arm and the entire camera system was left untouched during image acquisition, thereby preventing camera motion. (2) Images were taken from still scenes without any movements of objects induced by the operating surgeon. Therefore, motion artefacts could only originate from natural sources such as respiration and heartbeat and are thus not very strong and limited to images of thoracic organs (cf. example images in \autoref{fig:image_examples}).
Acquiring a fixed number of images with a fixed set of camera perspectives from a fixed set of situses per pig is infeasible in real-world surgery as no two surgeries are exactly the same. Tissues could be affected by a variety of complications that alter their state, such as inflammatory change or tissue trauma. Non-physiological tissues were excluded from the HSI acquisition, leading to variations in situses and number of images across pigs. The camera perspectives were chosen to allow for a good view of all organs of interest in the scene.

\begin{figure*}[htb]
    \centering
    \includegraphics[width=1\textwidth]{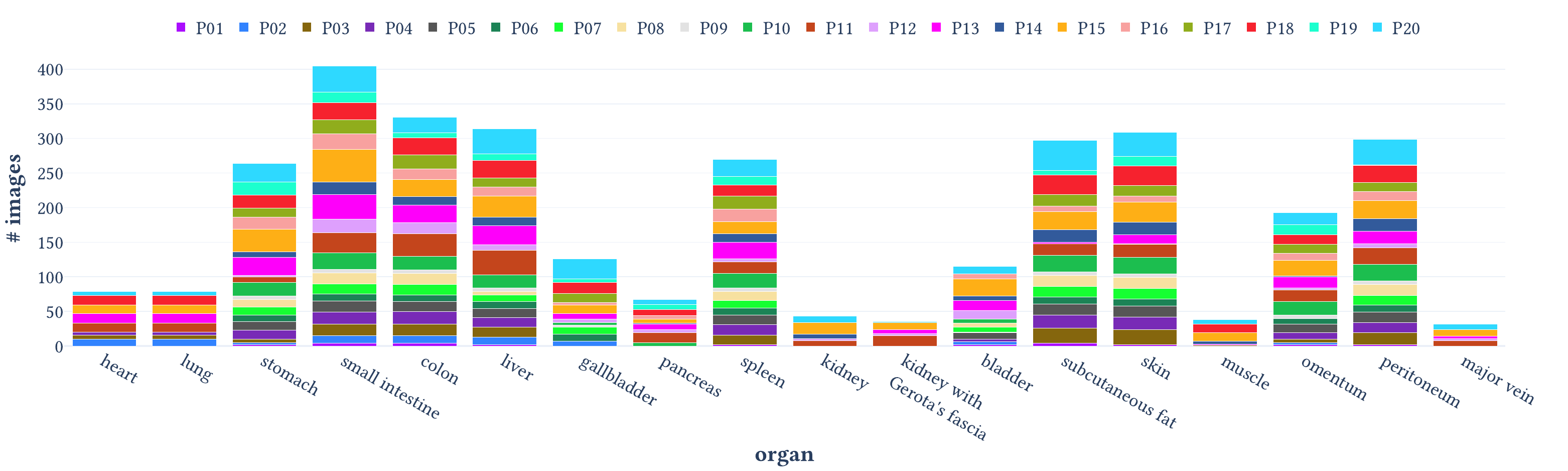}
    \caption[Data set overview]{Overview of the data set. \varTotalImages images from \varTotalPigs pigs have fully semantic annotations for \varTotalOrganClasses different organ classes and background. Each pig is represented by a unique color and denoted by \texttt{PXX}, where \texttt{XX} represents a unique pig identifier. Since background is naturally present on each image, the overview focuses on the number of images and pigs per organ.}
    \label{fig:dataset}
\end{figure*}

An overview of the data set is given in \autoref{fig:dataset}. In total, \varTotalImages images from \varTotalPigs pigs were acquired. For each organ, between \varMinImagesOrgan and \varMaxImagesOrgan images were acquired from \varMinPigsOrgan to \varMaxPigsOrgan individuals. To provide insights on characteristic organ spectra, an interactive figure is available in the supplementary materials. By selecting a pixel in an example image, the pixel spectrum as well as the mean spectrum and standard deviation across pixels of the corresponding organ class can be explored.

\subsubsection{Image annotation}

% suggestion for consistency: if referring to the naming of a label, put italic, if referring to the organ itself  (e.g. subcutaneous fat was annotated), don't put italic
\varTotalClasses different classes were annotated, including two thoracic organs (heart, lung), eight abdominal organs (stomach, small intestine, colon, liver, gallbladder, pancreas, kidney, spleen) and one pelvic organ (bladder). In the case of kidney, images before and after removal of Gerota's fascia were taken and labelled \oname{kidney with Gerota's fascia} and \oname{kidney}, respectively. Furthermore, subcutaneous fat, skin and muscle tissue as well as omentum, peritoneum and major veins were annotated. Pixels that belong to any inorganic object (e.g. cloth, compresses, foil, tubes, metallic objects and gloves) were labelled \oname{background}. This label is present on every image and the annotated areas cover on average \varBackgroundPixelRatio of an image. Additionally, pixels were labelled as \oname{ignore} if it could not be decided to which organ they belong or if they belonged to an organic object other than the \varTotalOrganClasses organ classes. \varInvalidImages of \varTotalImages images contain this label and, on average, the annotated areas cover \varInvalidPixelRatio of the pixels in the \varInvalidImages images. The \oname{ignore} pixels were later excluded from our analysis.

The semantic annotations were performed by two different annotators using vector annotation tools provided on the annotation platform SuperAnnotate (SuperAnnotate, Sunnyvale, USA)\footnote{\href{https://superannotate.com/}{https://superannotate.com/}}. To ensure consistent labelling, all annotations were then revised by the same medical expert.

Imbalances in the number of images per class arose since some organs naturally occur more often in the field of view of other organs. For example on images of the gallbladder, the surrounding liver is always present, whereas not on all liver images the gallbladder is visible. Heterogeneity in the number of animals per organ arose from differences in the surgical procedure performed. For example, opening of the thorax, which is a highly invasive and demanding surgical procedure and thus associated with a significant mortality and prolongation of the surgery, was only performed for eight out of \varTotalPigs pigs, making \ac{hsi} data from heart and lung unavailable for the remaining 12 pigs.

\subsection{Deep learning-based full semantic scene segmentation}
\label{sec:image_proc_pipeline}

Our approach to deep learning-based semantic \ac{hsi} image segmentation is summarized in \autoref{fig:model_overview}. The following sections present our image processing pipeline in detail, including an overview of our input modalities (\autoref{sec:modalities}), our pre-processing of the \ac{hsi} data (\autoref{sec:hsi_pre-proc}), the architectures of our deep-learning models (\autoref{sec:dl_models}), our training setup (\autoref{sec:training_setup}) and our approach to increase randomness in the data loading (\autoref{sec:data_loading}).

In general, our design choices with respect to model architectures and training setup were motivated by our comparative study. We aimed to have common model and training parameters across the different spatial models and modalities whenever possible (e.g. same hyperparameters and data splits). We explicitly avoided individual parameter tuning for each model to ensure a fair comparison and reduce computational costs.

\subsubsection{Model input modalities}
\label{sec:modalities}

A primary purpose of this study was to investigate whether there is a benefit in using \ac{hsi} data compared to RGB and \ac{tpi} data for neural network-based fully automated organ segmentation. For simplicity, we will refer to these different input data types as \textit{input modalities} although they were in practice all computed with the same camera. This reflects the fact that a future application leveraging semantic scene segmentation could be based on RGB images from a conventional camera, on the preprocessed \ac{hsi} images of an \ac{hsi} camera provider or on raw \ac{hsi} spectra. As illustrated in \autoref{fig:model_overview}, we trained neural networks separately on all three input modalities for all studied levels of data granularity. RGB data reconstructed from the \ac{hsi} data was available through the camera system. To study the organ segmentation performance on processed \ac{hsi} data, the associated \ac{sto2}, \ac{npi}, \ac{twi} and \ac{thi} images were stacked, yielding a $640\times480\times4$ ($w\times h\times c$) \ac{tpi} cube that served as model input.

\subsubsection{\ac{hsi} data pre-processing}
\label{sec:hsi_pre-proc}

In order to remove the influence of sensor noise and convert the acquired \ac{hsi} data from radiance to reflectance, the raw \ac{hsi} datacubes were automatically corrected with a pre-recorded white and dark reference cube by the camera system as described in \autocite{holmer_hyperspectral_2018}. After exporting the \ac{hsi} cubes from the camera system, the $\ell^1$-norm was applied to each pixel spectrum in order to account for multiplicative illumination changes that arise, for example, from fluctuations in the measurement distance.

\subsubsection{Deep learning models}
\label{sec:dl_models}

\begin{figure*}[htb]
    \centering
    \includegraphics[width=1\textwidth]{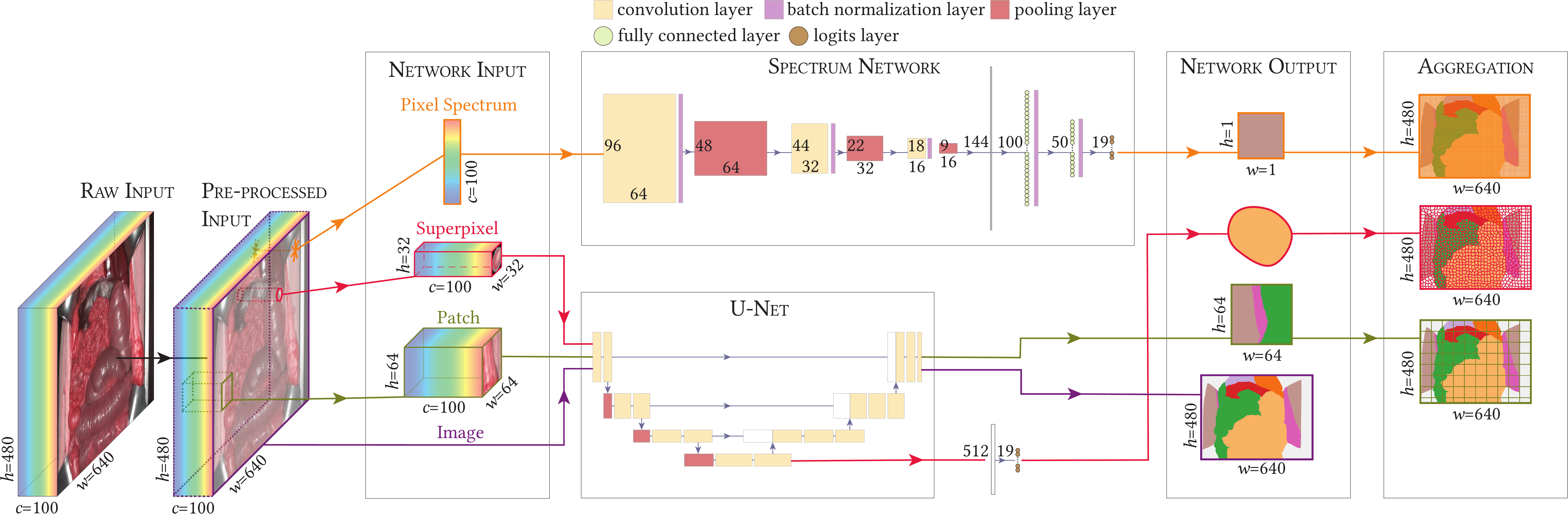}
    \caption[Model overview]{Overview of our deep learning pipeline for different spatial granularities based on \acf*{hsi} input data. For \textcolor{Pixel}{pixel-based classification}, pixel spectra are extracted from the pre-processed \ac{hsi} cubes and input to the spectrum network, yielding a pixel-wise class prediction. For \textcolor{Superpixel}{superpixel-based classification}, superpixel boundaries are segmented based on the reconstructed RGB images and each superpixel is then extracted from the pre-processed \ac{hsi} cubes. A minimum enclosing bounding box is computed and areas not belonging to the superpixel are replaced with zeros. After reshaping, the superpixel cube is input to the U-Net encoder and the output is further processed by a classification head. For \textcolor{PatchSixFour}{patch-based classification}, patches of a fixed shape are extracted from the pre-processed \ac{hsi} cubes and processed by the U-Net. Pixel-wise, superpixel-wise and patch-wise predictions belonging to the same image are aggregated to yield an image segmentation map. For \textcolor{Image}{image-based classification}, the input and output to the U-Net are a whole pre-processed \ac{hsi} cube and an image segmentation map, respectively.}
    \label{fig:model_overview}
\end{figure*}

Our approach to semantic scene segmentation is presented in \autoref{fig:model_overview} for the exemplary case of \ac{hsi} input data. Implementation details are provided in the following paragraphs.

\paragraph{Pixel-based segmentation}

The smallest possible spatial granularity of the input data is to use single pixel spectra, resulting in input feature vectors of length $c=100$ in the case of \ac{hsi} input data, $c=3$ in the case of RGB input data and $c=4$ in the case of \ac{tpi} input data.
Following our previous work in \autocite{studier-fischer_spectral_2021}, the deep learning architecture for the \ac{hsi} input data is composed of three one-dimensional convolutional layers with 64 filters in the first, 32 in the second, and 16 in the third layer. Each convolution is performed with a kernel size of 5 and operates on a single pixel spectrum (first layer) or across the spectral features from all filters of the previous layers (second and third layer). Each convolutional layer is followed by an average pooling layer with a kernel size of 2 which operates on the spectral features separately for each filter. The output of the last convolutional layer is stacked and serves as input for two fully connected layers, with 100 neurons in the first and 50 in the second layer. For \ac{tpi} and RGB input data, no convolutional operations along channels are feasible due to the small channel size. Instead, the network consists of three fully connected layers with 200 neurons in the first, 100 neurons in the second and 50 neurons in the third layer. The \ac{elu} \autocite{clevert_fast_2016} is used as an activation function and batch normalization is applied to the outputs of all layers except pooling layers. The class logits are calculated by a final linear layer, resulting in one organ label for each input pixel spectrum. The \ac{ce} loss function is used to optimize the model during training.

The architecture was designed such that local information from neighbouring spectral bands and global information across the entire spectrum is aggregated while keeping a small network size of only \varPixelHSITotalWeights weights for \ac{hsi}, \varPixelTPITotalWeights weights for \ac{tpi} and \varPixelRGBTotalWeights weights for RGB input data and thus being computationally efficient. The local information aggregation is achieved through the convolutional layers: a relatively small kernel size was chosen to focus on local structures and by stacking three convolutional layers, a compromise between increasing the receptive field of the network and keeping the number of learnable weights small was achieved \autocite{szegedy_rethinking_2016}. The global context was learned by the fully connected layers.

To retrieve a segmentation map for an image, we predict a class label for each pixel in the image and then map the resulting labels back to the image positions.

\paragraph{Superpixel-based segmentation}

Superpixels are regions of low spatial granularity that adhere to local boundaries which enclose pixels with similar features. As for the pixel-wise organ segmentation, the unsupervised clustering of superpixels turns the actual organ segmentation task into a superpixel-wise organ classification task. This is justified by the assumption that all pixels within a superpixel belong to the same organ class since superpixels are supposed to lie within the local boundaries of an organ. Superpixels are generated by the use of the simple linear iterative clustering (SLIC) algorithm on the reconstructed RGB data \autocite{achanta_SLIC_2012}. Prior to clustering, the image is smoothed with a Gaussian kernel of width 3 and then 1000 segments are computed in ten iterations while adaptively changing the per-superpixel compactness parameter (SLICO mode). For each superpixel, a minimum enclosing bounding box is computed and areas not belonging to the superpixel are replaced with zeros. To ensure one common input shape, superpixels are resized via bilinear interpolation to the shape $32 \times 32 \times 100$ ($w \times h \times c$) for \ac{hsi}, $32 \times 32 \times 4$ for \ac{tpi} and $32 \times 32 \times 3$ for RGB input data.

The resized superpixel cubes are passed to an efficientnet-b5 encoder \autocite{tan_efficientnet_2019} pre-trained on the ImageNet data set \autocite{deng_imagenet_2009} using the library of \citeauthor{yakubovskiy_segmentation_2020} \autocite{yakubovskiy_segmentation_2020}\footnote{\href{https://github.com/qubvel/segmentation_models.pytorch}{https://github.com/qubvel/segmentation\_models.pytorch}}. We chose this encoder as it yields good performance while at the same time being economical in terms of the number of parameters leading to a low memory footprint and fast computation times. The output of the encoder network is passed on to a classification head consisting of a fully connected layer with \varTotalClasses neurons for calculating the class logits. This way, the superpixel network shares the same architecture as the segmentation networks for the image and patch-based models with only minor modifications.

It is possible that not all pixels within one superpixel belong to the same organ class, for example due to inconsistencies at the border between organs. To account for this, we introduced the concept of fuzzy labels where we assigned a label vector of length $O$ to each superpixel (e.g. $O=19$ classes in our case). The fuzzy label vector stores the relative frequency of each class label considering all enclosed pixels inside the superpixel. The Kullback-Leibler divergence \autocite{kullback_information_1951} between fuzzy labels and the softmax output is used as a loss function during training.

During inference, the argmax of the class logits is computed and the resulting label is assigned to every pixel position of the superpixel. Predictions of all pixels from all superpixels are combined to yield a segmentation map for an image.

\paragraph{Patch-based segmentation}

Patches are regions of low spatial granularity that are extracted from images according to a fixed shape. They are generally more easily generated and more straightforward to use in neural networks than superpixels, for example because their rectangular shape matches with the rectangular kernel shapes of convolutional neural networks. In order to capture different degrees of granularity, patches of two different shapes are extracted: $32 \times 32 \times c$ and $64 \times 64 \times c$. These sizes serve as intermediate steps between the superpixel and the image model in terms of spatial granularity (cf. \autoref{tab:training_procedure}). We use patch sizes which are a power of two so as to easily integrate them with encoder architectures which halve the input shape multiple times. The number of generated patches per image corresponds to the number of patches that could have been generated via a grid-based tiling.

The patches are passed to a U-Net with an efficientnet-b5 encoder pre-trained on the ImageNet data set (like the superpixel network). Dice loss and \ac{ce} loss are calculated during training based on all pixels in the batch and equally weighted to compute the loss function. While each misclassified pixel contributes equally in the computation of the \ac{ce} loss, misclassified pixels belonging to an organ class of smaller image area contribute more to the dice loss than misclassified pixels belonging to a dominant class (e.g. \oname{background}). By computing a weighted sum of both loss terms, the network training can benefit from the respective advantages.

During inference, images are divided into a grid of non-overlapping patches of the corresponding patch size. In cases where an image dimension is not an integer multiple of the patch dimension, missing image regions are zero-padded. For each patch, the network yields a segmentation map. The segmentation maps of all patches of one image are combined to yield an image segmentation map. Segmentations belonging to previously zero-padded image regions are cropped.

\paragraph{Image-based segmentation}

Entire images offer a maximum of spatial granularity and we use them directly without any further adaptations of the image dimensions, i.e. the input tensors have a shape of $480 \times 640 \times 100$ ($w \times h \times c$) for \ac{hsi}, $480 \times 640 \times 4$ for \ac{tpi} and $480 \times 640 \times 3$ for RGB. Equivalent to the patch-based segmentation, the images are passed to an efficientnet-b5 U-Net pre-trained on the ImageNet data set. Again, both dice and \ac{ce} loss are equally weighted to compute the loss function.

\subsubsection{Training setup}
\label{sec:training_setup}

To prevent biases due to differences in the training setup when comparing the organ segmentation performance on input data for different levels of spatial granularity and modalities, the training setup was made as comparable as possible for all models. The following paragraphs describe the data augmentation, optimization and regularization methods that were consistently applied for all the different segmentation models.

\paragraph{Data augmentation}

In order to increase the size and diversity of the available training data and thereby improve convergence, generalization and robustness on out-of-distribution samples, data augmentation is commonly applied in computer vision \autocite{buslaev_albumentations_2020}. For all models, the training data is augmented using the Albumentation library\footnote{\href{https://github.com/albumentations-team/albumentations}{https://github.com/albumentations-team/albumentations}} with default settings \autocite{buslaev_albumentations_2020} on a per-image basis before extracting pixels, superpixels or patches: images are shifted (shift factor limit: 0.0625), scaled (scaling factor limit: 0.1), rotated (rotation angle limit: \SI{45}{\degree}) and flipped with a probability of $p$. All transformations are applied with a probability of $p=0.5$ to reduce the computational data loading costs that are induced by extensive data augmentations.

\paragraph{Optimization}

For all models, the Adam \autocite{kingma_adam_2017} optimization algorithm and an exponential learning rate scheme are used (initial learning rate: 0.001, decay rate $\gamma$: 0.99, Adam decay rates $\beta_1$: 0.9 and $\beta_2$: 0.999). Training is performed for 100 epochs and stochastic weight averaging \autocite{izmailov_averaging_2019} is applied. To achieve a comparable training procedure across models, the available training budget should be equal for each model. To this end, the size of one epoch is defined as seeing 500 training images for image-based segmentation. For pixel-, superpixel- and patch-based segmentation, it was ensured that the total number of extracted pixels for each approach matches roughly the total number of pixels of 500 images. Such matching can only be roughly approximated because the epoch size has to be divisible by the batch size so that every worker in the data loader can contribute equally to every batch, cf. \autoref{sec:data_loading}.

Recommendations for the optimal batch size are mixed (e.g. \autocite{smith_dont_2018, kandel_effect_2020}). While smaller batch sizes can, for example, speed up the learning process \autocite{montavon_neural_2012}, larger batches are a better representation of the real population, potentially leading to more stable gradients and better batch statistics \autocite{ioffe_batch_2015}. As a compromise, we decided to maximize the batch size while choosing a large number of epochs to rule out a potentially slower learning process. In practice, the batch size is constrained by the available \ac{gpu} memory and depends on the network and input size. We determined the maximum batch size thus per model, resulting in larger batch sizes for smaller input spatial granularities (cf. \autoref{sec:interpretation_results} for further discussion). \autoref{tab:training_procedure} gives an overview of the resulting epoch and batch sizes.

\begin{table}[!htp]
    \caption{Epoch and batch sizes for each of the different models. \#~pixels refers to the number of pixels of one input sample of a model. The model names patch\_64 and patch\_32 refer to models with the input shapes $32 \times 32 \times c$ and $64 \times 64 \times c$, respectively.}
    \label{tab:training_procedure}
    \centering
    \varEpochTable
\end{table}

During training, each model was evaluated after each epoch on the validation set of the respective training fold. The \ac{dsc} was calculated while respecting the hierarchical structure of the data, as described in more detail in \autoref{sec:experimental_conditions}. The final validation score was obtained by averaging the \ac{dsc} values of three validation pigs, and this score was also used to determine the best model across all epochs.

\paragraph{Regularization}

To avoid overfitting, dropout regularization is applied for the fully connected layers in the pixel models and in the superpixel classification head while using a dropout probability of 0.1.

\paragraph{Variability of results}
Training of neural networks is always subject to various sources of variation, some easier to control than others (e.g. seeding \textit{vs.} hardware influences) \autocite{pham_problems_2020}. Especially in our study, in which we aim for a fair comparison between different spatial granularities and modalities, reduction of such sources of variation is important. While obtaining exactly reproducible results is only achievable at the cost of longer training times (e.g. by restricting oneself to slow deterministic operations or a single homogeneous hardware infrastructure) \autocite{pham_problems_2020}, we took several measures to reduce the variation. We controlled the weight initialization of the networks, the initialization of the workers responsible for data loading (which also affects the data augmentation) and the ordering in which training samples are presented to the network for the modalities (e.g. corresponding spatial models for different modalities receive patches from the same spatial locations and in identical order). To this end, we set a seed for the network initialization and the data loaders and fixed the number of workers on each data loader to 12 across all experiments. Since we had to run many training runs efficiently for this study, we did not enforce deterministic operations and we used our in-house cluster infrastructure which consists of an inhomogeneous hardware infrastructure with many different GPUs (e.g. GeForce GTX\texttrademark{} 2080 Ti or DGX\texttrademark{} A100 (Nvidia Corporation, Santa Clara, USA)).

\subsubsection{Multi-image contribution in data loading}
\label{sec:data_loading}
\begin{figure}[htb]
    \centering
    \includegraphics[width=0.48\textwidth]{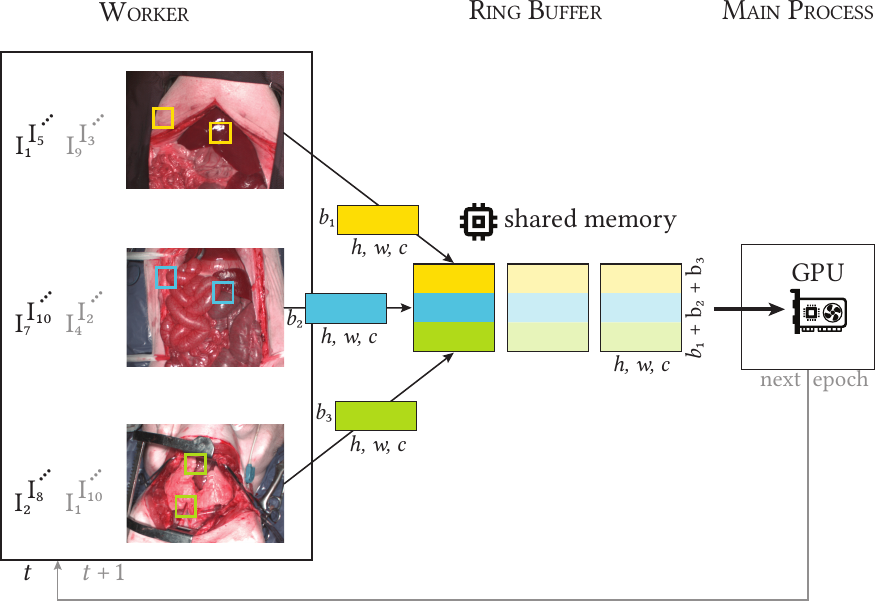}
    \caption[Data loading]{Data loading strategy designed to work with large data sets which ensures that batches are filled with image parts from multiple images. In this example, three workers generate random patches from their own set of images (mutually exclusive across workers), collect multiple patches in one batch part (boxes on the arrows) and store the result in the next free location in the ring buffer (non-transparent block) which is located in shared memory. The main process takes one of the next ready batches (transparent boxes) from the buffer and moves it to the \acf*{gpu}. $I_i$ refers to the $i$-th image in the data set, $b_j$ is the number of samples the $j$-th worker contributes to the batch and $h, w, c$ denote the height, width and number of channels of an image, respectively. The assignment of images $I_i$ to the workers changes randomly from one epoch $t$ to the next $t+1$.}
    \label{fig:data_loading}
\end{figure}

The models that work on part of an image such as the patch or pixel models feature an additional challenge when it comes to data loading. Due to the size of the data set, it is not feasible to load all images at the beginning of the training process as the \ac{hsi} data easily exceeds memory limitations. Pre-computing a set of image parts is also not feasible as the number of instances can be very high (e.g. each image has $480\cdot640=\num{307200}$ pixel instances). To overcome these issues, we used a custom data loading strategy outlined in \autoref{fig:data_loading} where all training images are distributed in unique sets to each worker. Each worker processes its own set of images as an endless stream but only ever loads one image at a time to extract the relevant parts (pixel, patch or superpixel in our case). The image parts are directly stored in a shared memory location by the workers so that the main process only needs to move the data to the \ac{gpu} without any further processing. The shared memory implements a ring buffer which gets filled by the workers in a cyclic manner where the next batch is filled as soon as the batch is moved to the \ac{gpu} by the main process. Every worker contributes equally to the batch, meaning that the same number of image parts are added to the batch dimension by every worker. This increases randomness across images in one batch so that the batch distribution more closely resembles the data distribution, which is usually desired when training a deep learning network. Once a worker has finished extracting parts from an image, the next image is loaded without keeping the old image in the memory. This leads to a constant memory footprint for data loading during training that is only determined by the number of used workers. The ordering of the images is reshuffled after each training epoch.

\section{Experiments and results}
\label{sec:experiments}

The purpose of our experiments was to investigate what constitutes an adequate representation of \ac{hsi} data for neural network-based organ segmentation (cf. \autoref{sec:hsi_representation}) and determine whether there is a benefit in using \ac{hsi} data over other modalities like RGB and \ac{tpi} data (cf. \autoref{sec:results_modalities}). The underlying experimental setup is described in \autoref{sec:experimental_conditions}.

\subsection{Experimental setup}
\label{sec:experimental_conditions}

We validated our framework on the data set described in \autoref{sec:data}. This section describes our train-test-split, validation metrics as well as how we assessed the quality of our reference annotations and how much training data was required.

\paragraph{Train-test-split} We split the data set comprising \varTotalPigs pigs (\varTotalImages images) into a train data set consisting of \varTotalTrainingPigs pigs (\varTotalTrainingImages images) and a disjoint test set consisting of five pigs (166 images). The five test pigs were randomly selected such that in both train and test set each organ class was represented by at least one pig. The test set remained untouched throughout model development. $k$-fold cross-validation was performed on the train data set with $k=5$. The folds were generated such that the number of organ classes was maximized across validation folds. We refer to this traditional validation set obtained for each fold, which consists of three pigs that are not seen during training, as $\mathcal{V}_\mathrm{unknown}$. Additionally, for each of the 12 pigs included in each of the five training sets, one random image was excluded from the training and used to compose a second validation set consisting of 12 unseen images per fold from pigs known from the training, referred to as $\mathcal{V}_\mathrm{known}$. By comparing the model's performance on the two validation sets, its generalization capabilities towards \varGeneralizationPhrasingS could be estimated. Such \varGeneralizationPhrasingS comprise changes due to inter-pig variability as well as imaging-related changes (e.g. different imaging perspectives) and context-related changes (e.g. variations in the performed surgery, surgical phase during acquisition and visible instruments). The latter two sources of variation may also be present across different images of the same surgery, but may be more prominent across different surgeries. Once model architectures and parameters were finalized based on the performance on the validation set, we evaluated the segmentation performance on the hold-out test data set by ensembling the network predictions for each fold via averaging the softmax values.

\paragraph{Validation metrics} Since individual validation metrics do not reflect all clinical requirements \autocite{yeghiazaryan_family_2018, reinke_common_2021}, we combined several validation metrics with different strengths and weaknesses to obtain a more informative quantification of the model performance. The most commonly used validation metric for biomedical segmentation tasks is the \ac{dsc} \autocite{dice_measures_1945, maier-hein_why_2018}. It measures the overlap between a predicted object segmentation and the corresponding reference segmentation and is highly sensitive to the object size and insensitive to the object shape \autocite{reinke_common_2021}. \ac{dsc} values are available per class and are always in the range $[0;1]$ with 0 indicating that either there is no overlap between predicted segmentation and reference segmentation for the class or the class is present on the image but has not been predicted. A value of 1 for a class is obtained if the prediction overlaps perfectly with the reference segmentation. It should be noted that any binary segmentation task can be regarded as a pixel-wise classification tasks and that the \ac{dsc} is identical to the F1-score, which is typically represented as a function of true/false positives/negatives. Furthermore, the \ac{dsc} is very closely related to the Intersection over Union (IoU), which is commonly used as primary validation metric in the machine learning community. For the computation of the \ac{dsc}, we used the MONAI framework\footnote{\href{https://monai.io}{https://monai.io}} \autocite{consortium_monai_2020}.

Boundary-distance-based methods measure the dissimilarity between the predicted segmentation and reference segmentation in terms of distances between boundaries. In contrast to overlap-based metrics, boundary-distance-based metrics are insensitive to the object size and sensitive to the object shape. An example is the \ac{asd} \autocite{heimann_comparison_2009} which for an organ $o$ is the average of all distances between pixels on the predicted object segmentation border $\mathcal{B}_{\mathrm{ML}}^o$ and its nearest neighbour on the reference segmentation border $\mathcal{B}_{\mathrm{REF}}^o$. Here, we used the symmetric version of the \ac{asd} which repeats the computation of the set of nearest neighbour distances $\mathcal{D}_{\mathrm{ML}}^o$ with the role of the predicted segmentation and reference reversed, yielding $\mathcal{D}_{\mathrm{REF}}^o$. All obtained distance values are averaged, yielding an average distance value $\ac{asd}^o$ for each class:

\begin{equation}
    \ac{asd}^o = \frac{\sum\limits_{d \in \mathcal{D}_{\mathrm{ML}}^o} d + \sum\limits_{d^{'} \in \mathcal{D}_{\mathrm{REF}}^o} d^{'} }{\left|\mathcal{D}_{\mathrm{ML}}^o\right| + \left|\mathcal{D}_{\mathrm{REF}}^o\right|}
\end{equation}

The \ac{asd} has the disadvantage that it is unbounded, yielding values in the range $[0;\infty)$ with 0 in cases where boundaries of objects match exactly. Therefore, \ac{asd} values are generally harder to interpret. Furthermore, special attention must be placed on the case of missed classes (classes that are present in the reference annotations but were not predicted), as no natural limit is available \autocite{reinke_common_2021}. Here, we decided to set the \ac{asd} value for a missed class to the maximum \ac{asd} obtained for the other classes on the same image. This introduced a potentially high and image-dependent penalty when a class could not be predicted in an image (cf. discussion on this point in \autoref{sec:limitations}).

The \ac{nsd} \autocite{nikolov_deep_2021} estimates which fraction of a segmentation boundary is correctly predicted with an additional threshold $\tau$ related to the clinically acceptable amount of deviation in pixels. It is thus a measure of what fraction of a segmentation boundary would have to be redrawn to correct for segmentation errors. Instead of one common threshold $\tau$, we used a class-specific threshold $\tau^o$ for each organ class $o$ since the difficulty of annotating varies between organs (e.g. an organ with a clear boundary, such as liver, is easier to precisely annotate than an organ with a fuzzy boundary, such as omentum). We adapted the \ac{nsd}, which was originally invented for 3D segmentation maps, to our 2D segmentation maps in the following way: Instead of considering 3D segmentation surfaces, we considered 2D segmentation boundaries. For all pixels of the predicted segmentation boundary of organ $o$, $\mathcal{B}_{\mathrm{ML}}^o$, the nearest-neighbour distances to the reference segmentation boundary $\mathcal{B}_{\mathrm{REF}}^o$ were computed, resulting in a set of distances $\mathcal{D}_{\mathrm{ML}}^o$. We determined the subset $\mathcal{D}_{\mathrm{ML}}^{'o}$ of distances in $\mathcal{D}_{\mathrm{ML}}^o$ that are smaller or equal to the acceptable deviation $\tau^o$
\begin{equation}
    \mathcal{D}_{\mathrm{ML}}^{'o} = \{ d \in \mathcal{D}_{\mathrm{ML}}^{o} \, | \, d \leq \tau^o \}.
\end{equation}
The entire procedure was symmetrically repeated for $\mathcal{B}_{\mathrm{REF}}^o$, yielding $\mathcal{D}_{\mathrm{REF}}^o$ and $\mathcal{D}_{\mathrm{REF}}^{'o}$. For each class that appears in both the prediction and the reference segmentation the $\ac{nsd}^o$ was then computed as:

\begin{equation}
    \ac{nsd}^o = \frac{\left|\mathcal{D}_{\mathrm{ML}}^{'o}\right| + \left| \mathcal{D}_{\mathrm{REF}}^{'o} \right|}{\left|\mathcal{D}_{\mathrm{ML}}^o\right| + \left|\mathcal{D}_{\mathrm{REF}}^o\right|}
\end{equation}

The \ac{nsd} is bounded in the range $[0;1]$ with 0 indicating that either the boundary is completely off, with all distances being larger than the acceptable deviation $\tau^o$, or that the class is present on the image, but has not been predicted. 1 is obtained for cases where no redrawing of the segmentation boundary is necessary since every deviation is below the acceptable threshold $\tau^o$.

One of the major challenges of the \ac{nsd} is the determination of the organ-specific thresholds $\tau^o$ (cf. discussion in \autoref{sec:limitations}). To this end, \varRaterTotalImages randomly selected images (one image per pig such that all organ classes are represented by at least two images) were re-annotated by a second medical expert. Equivalent to the \ac{asd}, we computed distances between the boundaries of the original and the re-annotation for each organ $o$ in each image $i$ and averaged the results to obtain the image- and organ-specific threshold $\tau_i^o$. If an organ was annotated in only one of the two corresponding images, no distances could be computed and the corresponding structure was ignored (cf. discussion in \autoref{sec:limitations}). We computed the class-specific distance threshold $\tau^o$ by averaging the $\tau_i^o$ for the set of images $\mathcal{I}^o$ where the organ $o$ is present and $\tau_i^o$ could be computed:

\begin{equation}
    \label{eq:nsd_thresholds}
    \tau^o = \frac{1}{\left|\mathcal{I}^o\right|} \sum_{i \,\in\, \mathcal{I}^o} \tau_i^o.
\end{equation}

For each image, metric values were first separately computed for all classes annotated in the reference segmentation. These class-wise metric values were then averaged to obtain a per-image metric value. To account for the hierarchical nature of the data (following \autocite{holland-letz_drawing_2020}), all metric values were first averaged for all images of one pig and these per-pig scores served as a basis for visualizations and model rankings.

We investigated the model ranking and its stability with respect to two sources of variability, namely the sampling variability and variability due to choice of validation metric. The ranking analyses were performed using the challengeR\footnote{\href{https://github.com/wiesenfa/challengeR}{https://github.com/wiesenfa/challengeR}} toolkit \autocite{wiesenfarth_methods_2021}. 

The model ranking was determined in the following way: For each model, the average of the five per-pig metric values was computed and models were ranked according to these mean metric values. To assess the stability of the ranking with respect to different validation metrics, the ranking was performed separately for each metric and compared. To assess the stability of the model ranking concerning the sampling variability, rankings were performed repeatedly on \num{1000} bootstrap samples. Each bootstrap sample consisted of 5 cases randomly drawn with replacement from the set of five test cases (one metric value for each pig in the test set). Metric values were first averaged across all five test cases and then the rank of each model was determined based on this aggregate, resulting in \num{1000} ranks for each model (one rank per bootstrap sample, \texttt{meanThenRank} option in challengeR).

\paragraph{Quality of reference annotations} Crucial for any deep learning algorithm is the quality of the data, including the available reference annotations. It is known from previous studies that the variability between different human raters can be large \autocite{joskowicz_inter-observer_2019}. To quantify the quality of our reference annotations, we used the same set of re-annotated images as described above and the annotations of the second medical expert for inter-rater estimations. Additionally, the medical expert who originally annotated our data set re-annotated the same set of \varRaterTotalImages images for an estimation of the intra-rater variability. In both cases, we compared the re-annotations with the original annotation per image and calculated our three evaluation metrics. In contrast to the comparison of model predictions, we did not use pixels which the new annotator assigned to the \oname{ignore} class (e.g. pixels for which an annotator was not sure). That is, the union of ignored pixels in reference and re-annotated segmentation map was computed and those pixels were ignored.

\paragraph{Required amount of training data}
To study the effect of the amount of training data on the performance of the different models, we randomly sampled $n$ pigs from the set of \varTotalTrainingPigs training pigs without replacement and varied $n$ from 1 to \varMaxTrainingPigsDatasetSize. Training of the models was only performed on the images of the $n$ sampled pigs without $k$-folds, while the performance was evaluated on the same test set described in \autoref{sec:experimental_conditions} but only for \varTotalClassesDatasetSize organ classes and without ensembling. These \varTotalClassesDatasetSize organ classes (\varClassesDatasetSize) are the set of organs of which images are available for all \varTotalTrainingPigs training pigs. This design choice avoids the problem of having a pig sampled during training that does not contain any of the target classes, and hence would not give indicative results for this experiment. To increase the stability of the results towards pig sampling variability, we repeated the experiment five times with different random pig selections.

\paragraph{Misclassifications}
Misclassifications were analysed by computing the confusion matrix for the model of interest. To account for the hierarchical structure of the data, confusion matrices were first computed per pig and row-wise $\ell^1$-normalized by dividing every value by the total number of pixels of the respective organ in the reference segmentation. Thereupon, the row-normalized confusion matrices of all test set pigs were averaged to yield the model-wise confusion matrix.

\subsection{Representation of \ac{hsi} data}
\label{sec:hsi_representation}

A primary purpose of our study was to investigate what is an adequate representation of \ac{hsi} data both with respect to the segmentation performance, the required amount of training data and the capability to generalize towards \varGeneralizationPhrasingS. The experiments performed to answer these research questions are outlined together with their results in the following paragraphs.

\paragraph{Model performance} \autoref{fig:benchmarking_box} shows the performance of the pixel-based (referred to as pixel), superpixel-based (referred to as superpixel), patch-based (referred to as patch\_32 for an input shape of $32 \times 32 \times c$ and patch\_64 for an input shape of $64 \times 64 \times c$) and image-based (referred to as image) segmentation models for our three validation metrics \ac{dsc}, \ac{asd} and \ac{nsd}. Despite the fact that performance gaps for different spatial granularities are less prominent for \ac{hsi} data than for RGB and \ac{tpi} data, a consistent trend is visible for all levels of spectral granularity and validation metrics: the larger the spatial granularity of the input data, the better the segmentation performance. 

\newcommand{\benchmarkingBoxDescription}[1]{Segmentation performance of the spatio-spectral models. The performance of five spatial models and three modalities (RGB, \acf*{tpi} and \acf*{hsi}) is shown for three different metrics on the #1 set. The dotted line and the shaded area represent the mean and standard deviation range, respectively, of the inter-rater results for the corresponding metric. Each boxplot shows the quartiles of the metric value distribution with the whiskers extending up to $1.5$ times the interquartile range, and the median and mean as solid and dashed line, respectively. Each dot represents one test pig.}
\begin{figure}[h!]
\begin{subfigure}{0.48\textwidth}
  \centering
  \includegraphics[width=1\linewidth]{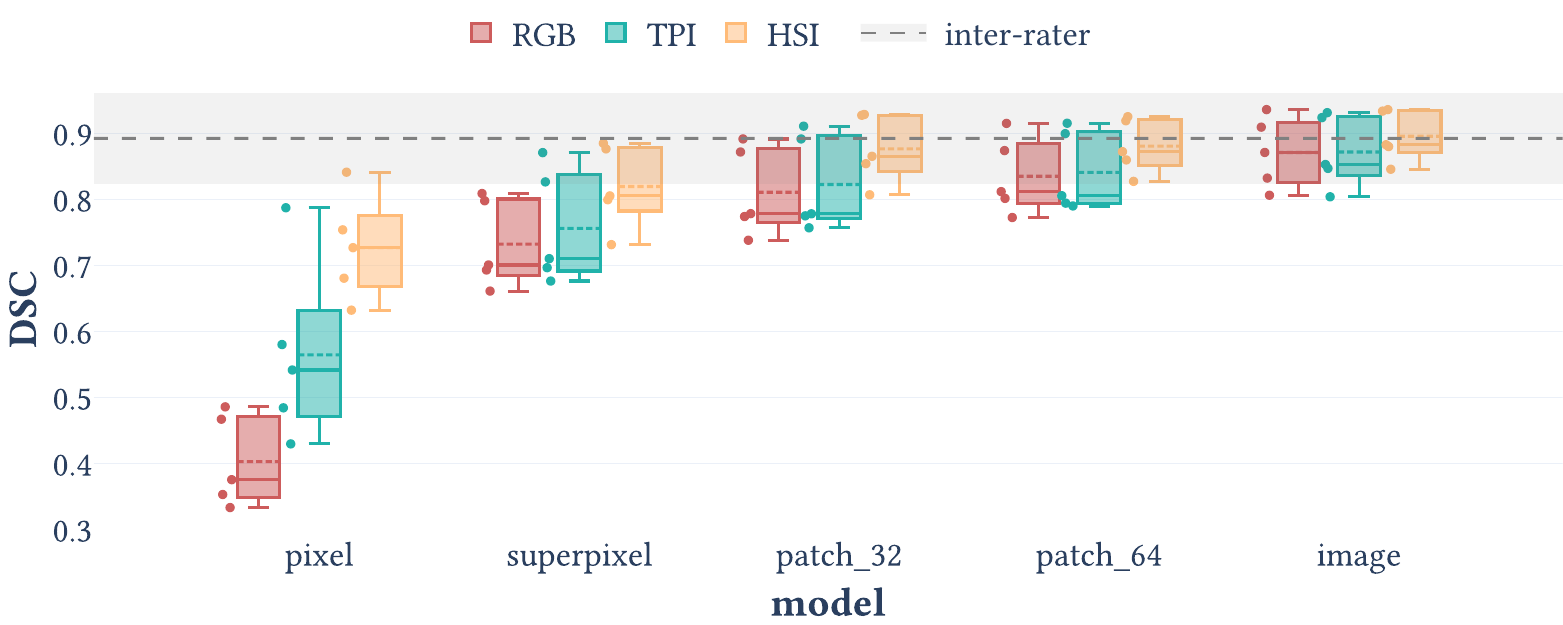}
  \caption{\Acf*{dsc}}
  \label{fig:benchmarking_box_DSC}
\end{subfigure}
\begin{subfigure}{0.48\textwidth}
  \centering
  \includegraphics[width=1\linewidth]{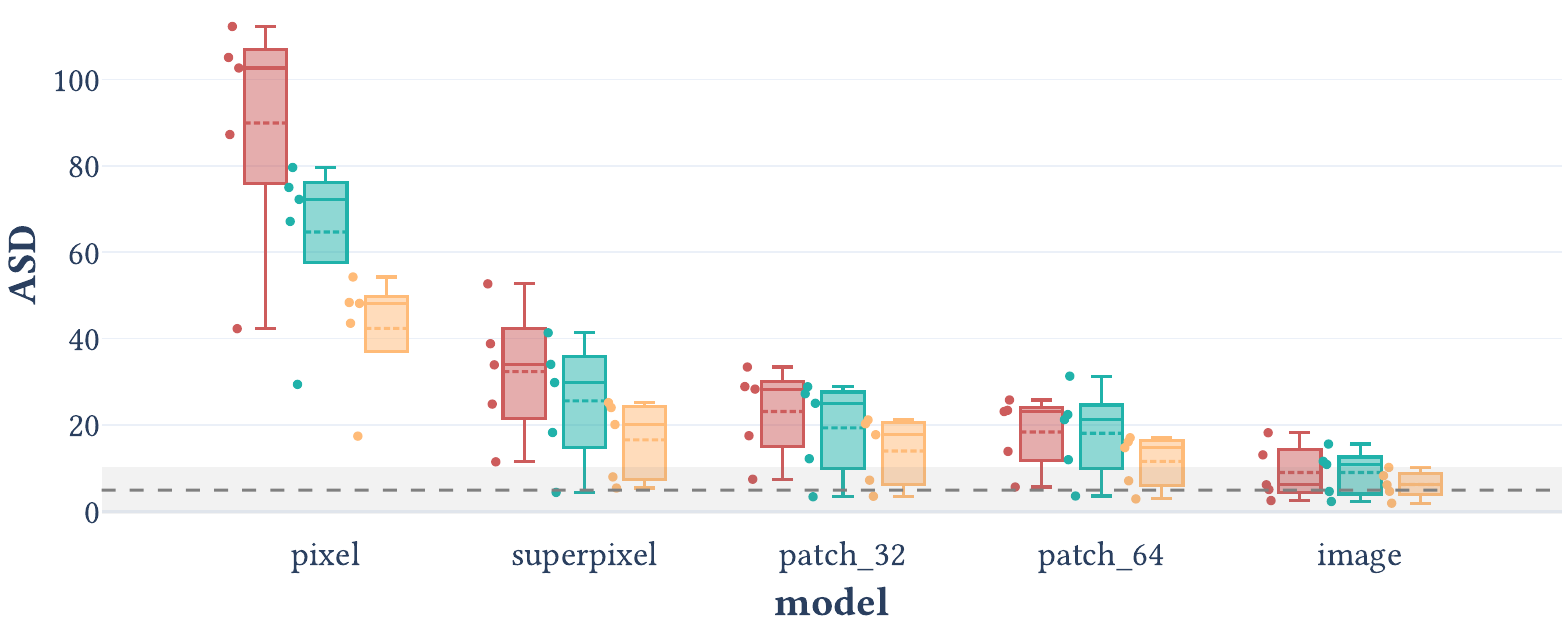}
  \caption{\Acf*{asd}}
  \label{fig:benchmarking_box_ASD}
\end{subfigure}
\begin{subfigure}{0.48\textwidth}
  \centering
  \includegraphics[width=1\linewidth]{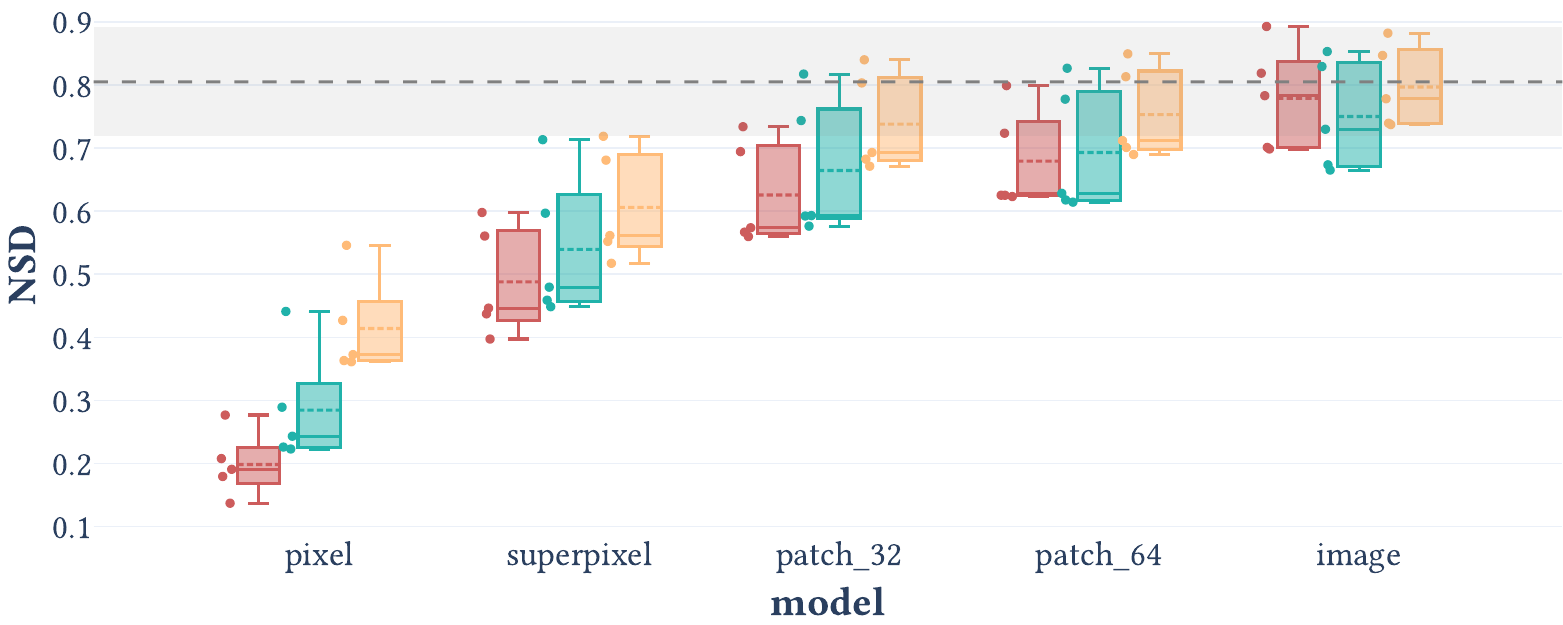}
  \caption{\Acf*{nsd}}
  \label{fig:benchmarking_box_NSD}
\end{subfigure}
\caption{\benchmarkingBoxDescription{test} The results on the validation data set reveal a similar pattern and are shown in \autoref{fig:benchmarking_box_val}.}
\label{fig:benchmarking_box}
\end{figure}

Notably, the image-based \ac{hsi} organ segmentation model performance is, on average, consistently comparable to predictions of a second human expert for all validation metrics. For the inter-rater variability, we obtained a \ac{dsc} of \varRaterInterDSC, an \ac{asd} of \varRaterInterASD and an \ac{nsd} of \varRaterInterNSD. The intra-rater variability is better on all three metrics with a \ac{dsc} of \varRaterIntraDSC, an \ac{asd} of \varRaterIntraASD and an \ac{nsd} of \varRaterIntraNSD. Across all \varRaterTotalImages images, it occurred \varRaterInterAdditional times in the inter-rater and \varRaterIntraAdditional times in the intra-rater agreement evaluation that organ classes that were not annotated in the reference segmentation map were newly assigned to an image. \varRaterInterMissing times in the inter-rater and \varRaterIntraMissing times in the intra-rater agreement evaluation, organ classes that were annotated in the reference segmentation map were missing in the re-annotations. Differences in the \oname{ignore} class occurred for \varRaterInterTotalMaskDiffImages and \varRaterIntraTotalMaskDiffImages out of the \varRaterTotalImages images in the inter-rater and intra-rater comparison, respectively. In total, for \varRaterInterTotalMaskDiffPixels in the inter-rater and \varRaterIntraTotalMaskDiffPixels in the intra-rater comparison, the label \oname{ignore} was assigned in the re-annotation, but an organ label had been assigned in the reference annotation or vice-versa.

We described our strategy for the reduction of software- and hardware-related variability of our results in \autoref{sec:training_setup}. To get an estimate of the controlled source of variation, we ran the image-based \ac{hsi} model 5 times with different seeds and found the \ac{dsc} to be in the range $[\texttt{min}; \texttt{max}] = \varSeedVariationTestDSC$, the \ac{asd} in $\varSeedVariationTestASD$ and the \ac{nsd} in $\varSeedVariationTestNSD$ on the test set. The controlled software- and hardware-related variability between pigs in the test set is thus at least one magnitude smaller than the inter-pig variability.

\paragraph{Model ranking}
In \autoref{fig:ranking_dsc}, we illustrate the ranking stability with respect to the sampling variability for the \ac{dsc} (results for the \ac{asd} and \ac{nsd} can be found in \autoref{fig:ranking_asd} and \autoref{fig:ranking_nsd}, respectively). The bootstrapped ranking is relatively stable with at least the first and last two ranks being very clear (more than \SI{90}{\%} of bootstraps resulting in the same rank) for all metrics. For the boundary-distance-based metrics, the number of models with a clear ranking is even larger and for the \ac{nsd}, all ranks vary by a maximum of $\pm 1$ rank around the median.

\newcommand{\rankingDescription}[3]{Ranking stability of the different models and modalities based on bootstrap sampling on the test set with the \acf*{#1}. The area of each blob at position ($A_i$, rank $j$) is proportional to the relative frequency of algorithm $A_i$ (model\#modality combination) achieving rank $j$ across \num{1000} bootstrap samples with one bootstrap sample consisting of 5 pig-level metric values. The median rank for each algorithm is indicated by a black cross and the black lines indicate the \SI{95}{\percent} quantile of the bootstrap results. This plot was generated with the challengeR toolkit \autocite{wiesenfarth_methods_2021}. Ranking results for the \acf*{#2} and \acf*{#3} can be found in \autoref{fig:ranking_#2} and \autoref{fig:ranking_#3}, respectively.}
\begin{figure}[htb]
    \centering
    \includegraphics[width=0.48\textwidth]{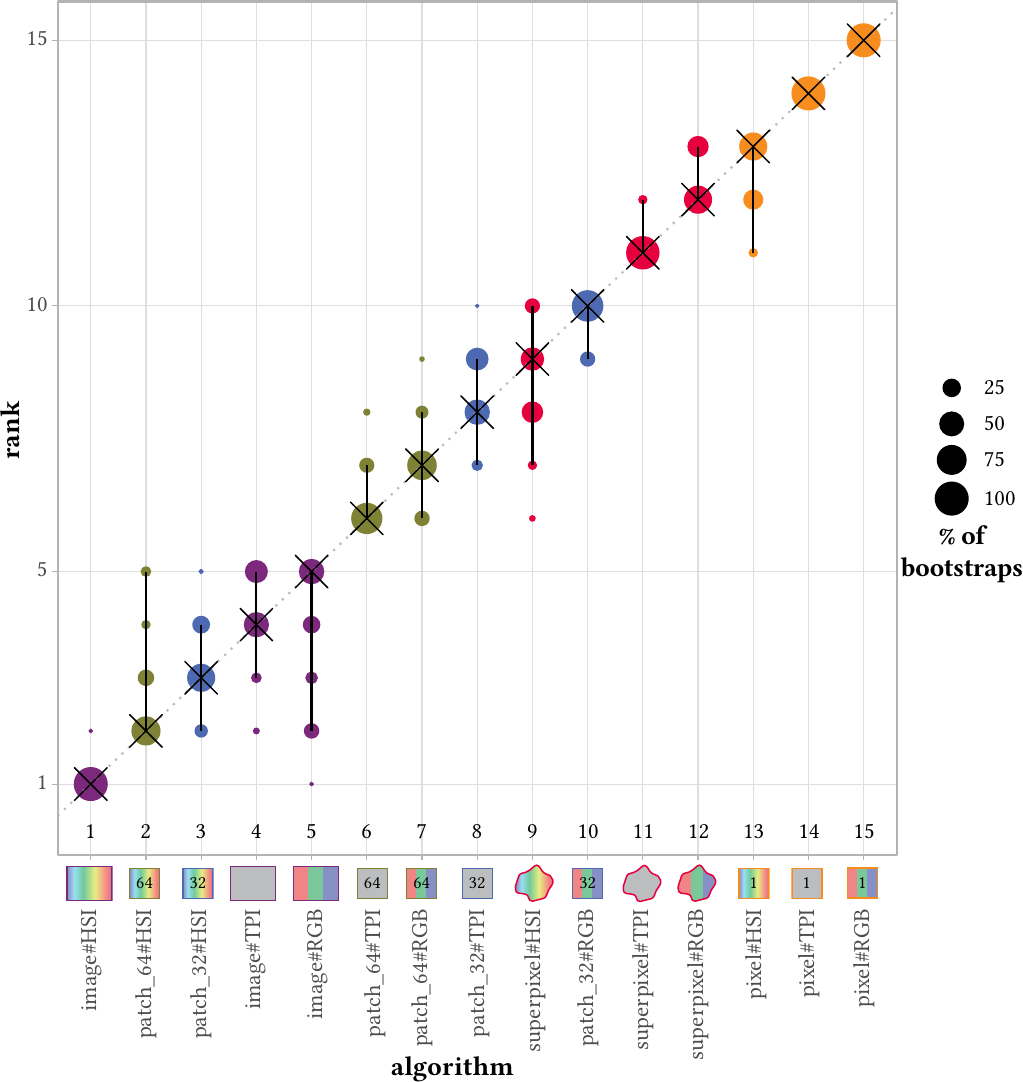}
    \caption[Ranking]{\rankingDescription{dsc}{asd}{nsd}}
    \label{fig:ranking_dsc}
\end{figure}

\begin{figure}[htb]
    \centering
    \includegraphics[width=0.48\textwidth]{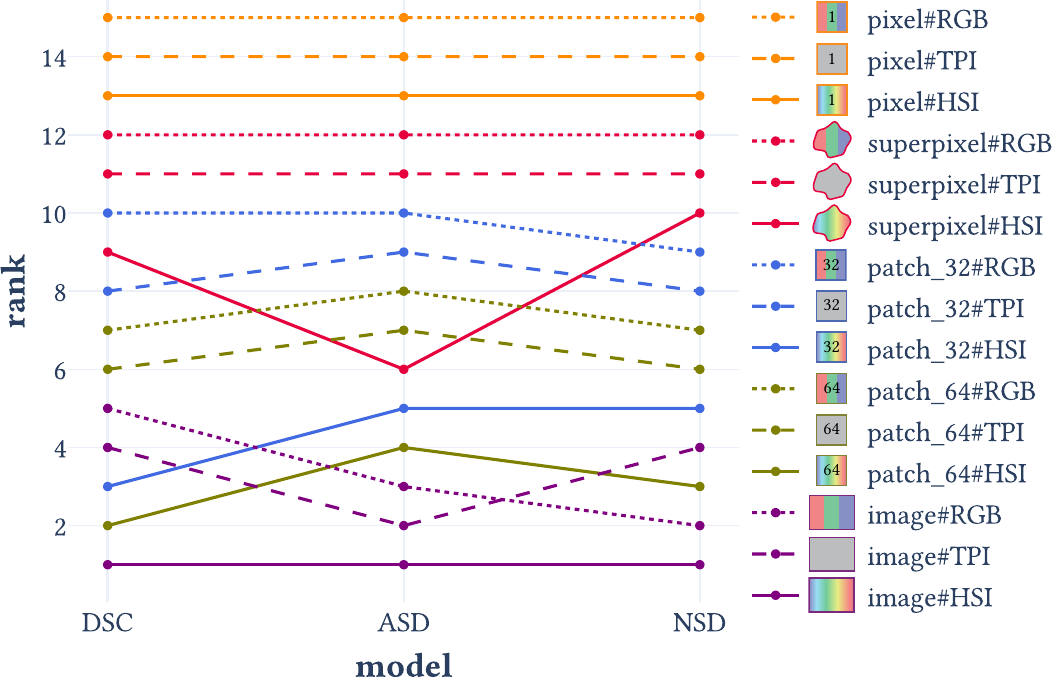}
    \caption[Ranking Metrics]{Ranking stability of the different models and modalities (RGB, \acf*{tpi} and \acf*{hsi}) for three different metrics (\acf*{dsc}, \acf*{asd} and \acf*{nsd}) on the test set. Each line visualizes how the rank of the corresponding model\#modality changes when evaluating with different metrics. This plot is based on ranking data from the challengeR toolkit \autocite{wiesenfarth_methods_2021}.}
    \label{fig:ranking_metrics}
\end{figure}

To assess the stability of the ranking with respect to different validation metrics, rankings for the three metrics were compared as depicted in \autoref{fig:ranking_metrics}. Across all modalities and metrics, the ranking of the spatial models is always (from best to worst): image, patch\_64, patch\_32, superpixel and pixel. Hence, more context always improves the segmentation performance irrespective of the modality and metric. Generally, rankings for the different metrics are in close agreement: The image-based segmentation of \ac{hsi} data always ranks first, while the last five ranks are always taken by superpixel\#\ac{tpi}, superpixel\#RGB, pixel\#\ac{hsi}, pixel\#\ac{tpi} and pixel\#RGB (from best to worst). The largest difference in ranking across metrics occurs for the superpixel\#\ac{hsi} model, which achieves rank six for the \ac{asd} compared to rank nine and ten for the \ac{dsc} and \ac{nsd}, respectively.

\begin{figure*}[h!]
    \centering
    \includegraphics[width=\textwidth]{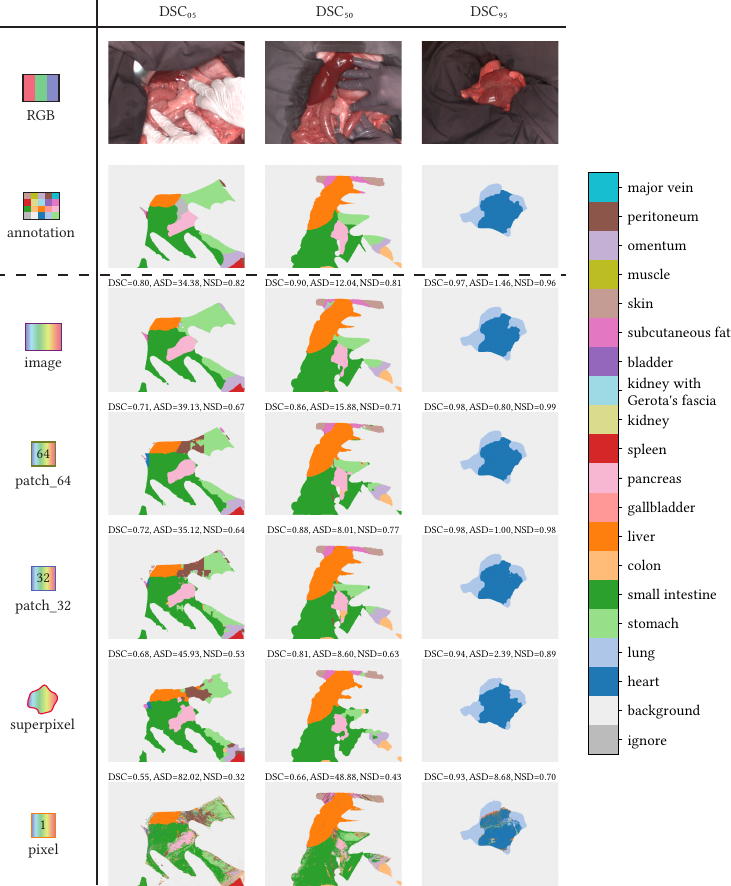}
    \caption[Image examples]{Example predictions for the different spatial models on the \acf*{hsi} modality. Images are selected based on the $q$\,\si{\percent} quantile of the \acl*{dsc} averaged across all five models ($\text{DSC}_{q}$). \oname{ignore} summarizes instances which are excluded in our models (e.g. pixels with \oname{unsure} label). For each prediction, metric values for the \ac{dsc}, \acf*{asd} and \acf*{nsd} are shown.}
    \label{fig:image_examples}
\end{figure*}

\paragraph{Visual segmentation quality assessment} Example predictions for the five spatial models on the \ac{hsi} data are shown in \autoref{fig:image_examples}. Based on the image \ac{dsc} averaged over all five models, the images corresponding to the \SI{5}{\%} quantile, \SI{50}{\%} quantile and \SI{95}{\%} quantile were selected, representing examples for overall bad, intermediate and good segmentation performances, respectively. For pixel-based segmentation predictions, boundaries are more incomplete and scattered than for the other models. In some of the patch-based segmentation examples, sharp vertical and horizontal edges are visible where adjacent patches connect, whereas, for the superpixel-based segmentation, edges appear wiggly due to misclassified superpixels in the proximity of organ boundaries.

\paragraph{Required amount of training data} A potential benefit when using input data of smaller spatial granularity is that more training samples are available, e.g. one image corresponds to \num{307200} training samples in the case of pixel-based segmentation, but only to a single training sample in the case of image-based segmentation (\autoref{tab:training_procedure}). \autoref{fig:dataset_size_dsc} shows the development of the \ac{dsc} over the number of training pigs for the different models. Results for \ac{asd} and \ac{nsd} can be found in \autoref{fig:dataset_size_asd} and \autoref{fig:dataset_size_nsd}, respectively. For all studied numbers of training pigs, the performance of image-based segmentation on \ac{hsi} is comparable or better than the performance of other models.

\newcommand{\datasetSizeDescription}[3]{\Acf*{#1} performance on the test set as a function of the number of individuals $n$ in the training set for different models. The solid line represents the average performance and the shaded area 0.5 standard deviations across 5 runs with a different random selection of individuals in the training set. Only results for the \acf*{hsi} modality are shown. Results for the \acf*{#2} and \acf*{#3} are shown in \autoref{fig:dataset_size_#2} and \autoref{fig:dataset_size_#3}, respectively.}
\begin{figure}[htb]
    \centering
    \includegraphics[width=0.48\textwidth]{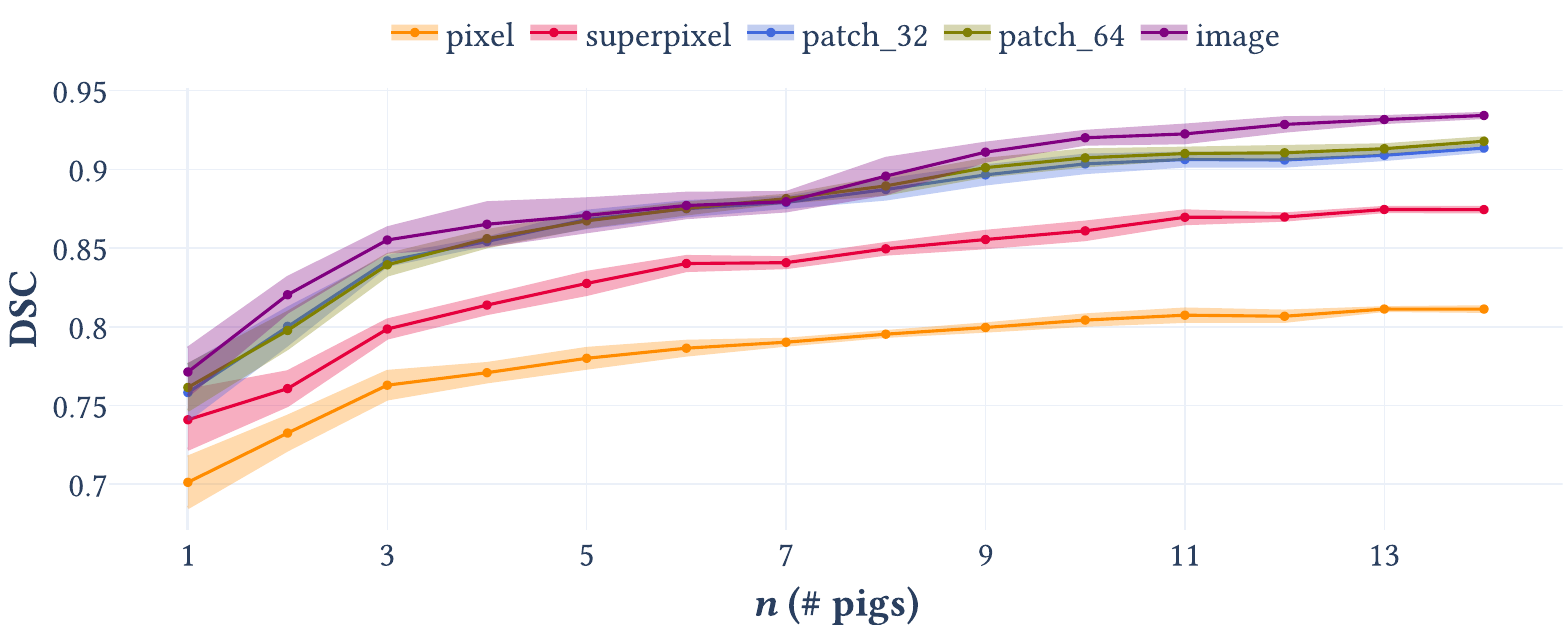}
    \caption[Data set size]{\datasetSizeDescription{dsc}{asd}{nsd}}
    \label{fig:dataset_size_dsc}
\end{figure}

\begin{figure}[ht]
\begin{subfigure}{0.48\textwidth}
  \centering
  \includegraphics[width=1\linewidth]{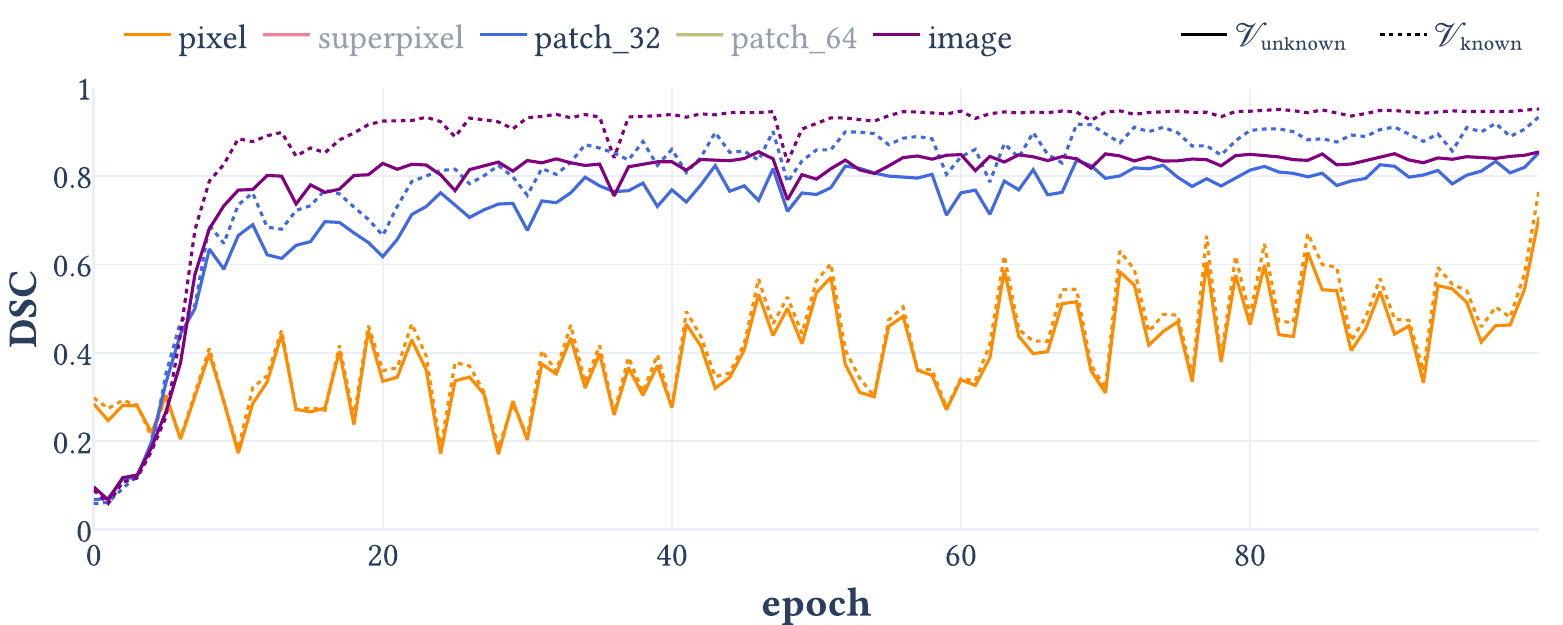}
  \caption{\Acf*{hsi}}
  \label{fig:generalization_error_hsi}
\end{subfigure}
\begin{subfigure}{0.48\textwidth}
  \centering
  \includegraphics[width=1\linewidth]{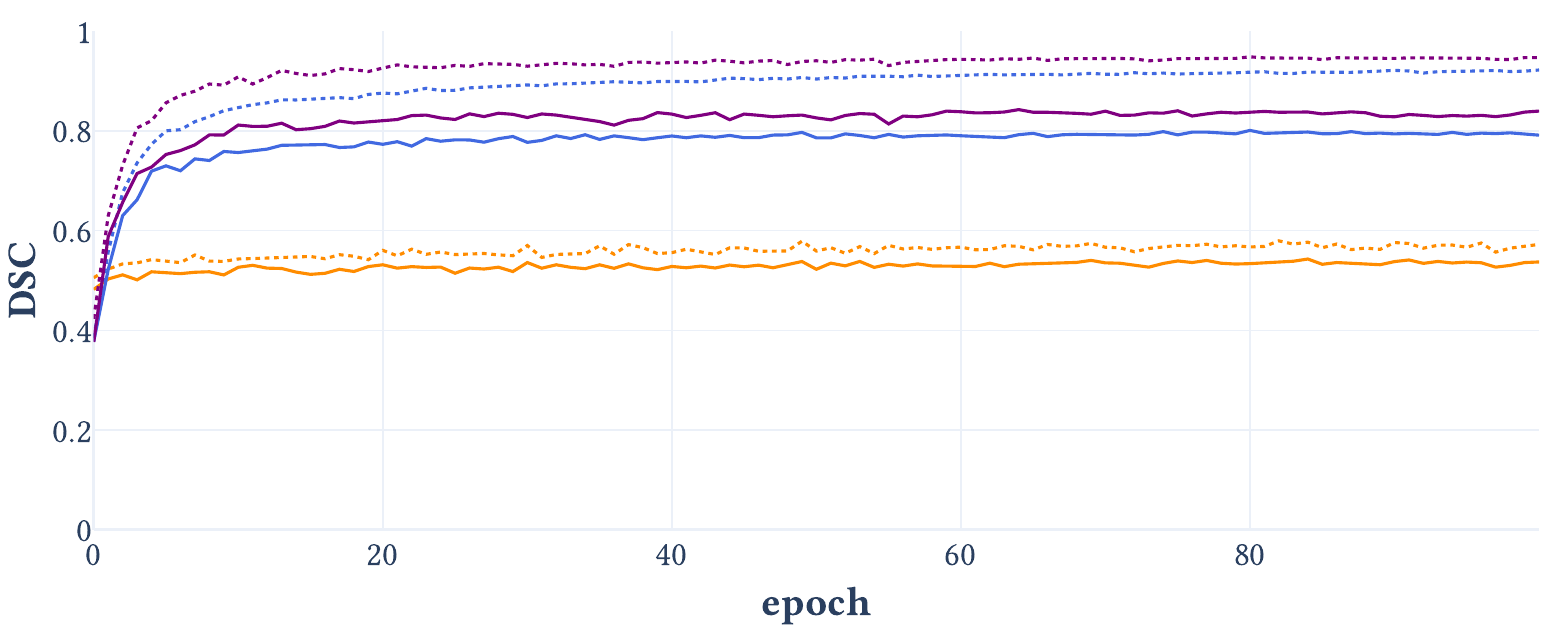}
  \caption{\Acf*{tpi}}
  \label{fig:generalization_error_param}
\end{subfigure}
\begin{subfigure}{0.48\textwidth}
  \centering
  \includegraphics[width=1\linewidth]{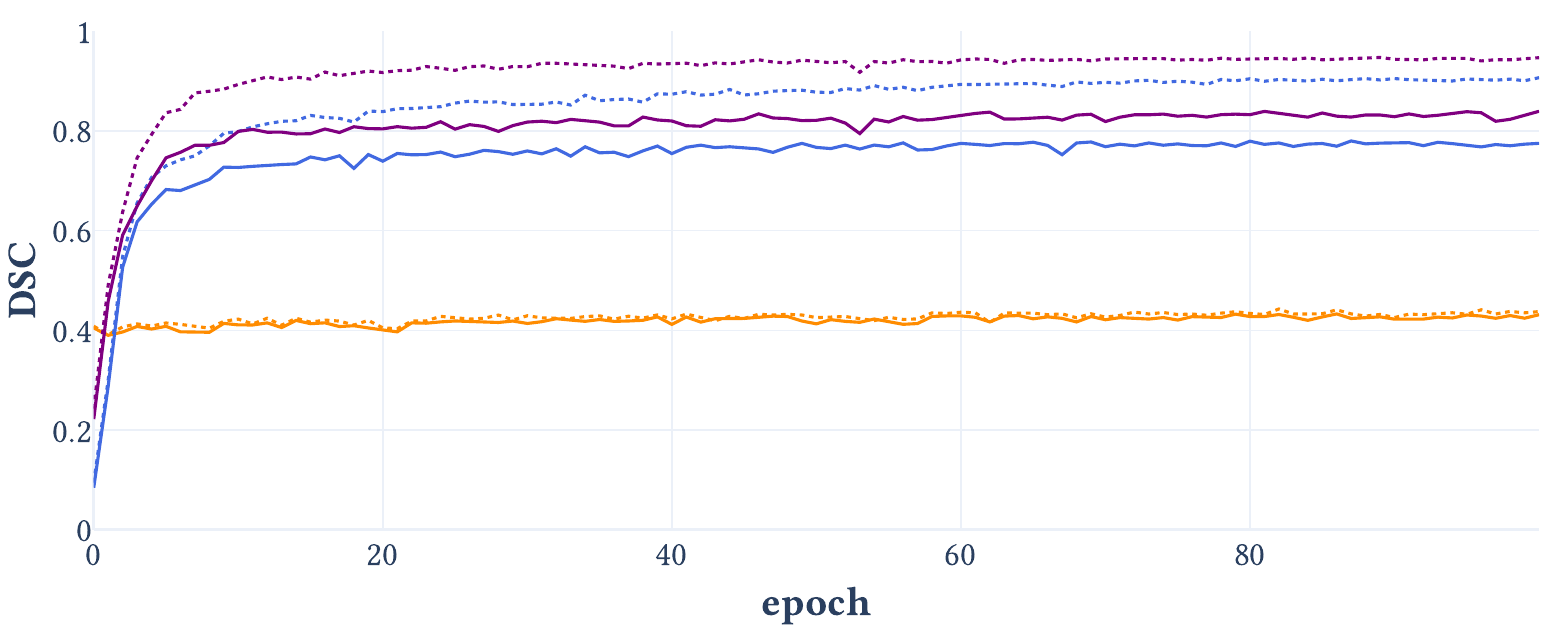}
  \caption{RGB}
  \label{fig:generalization_error_rgb}
\end{subfigure}
\caption{Generalization error over training time by comparing the two validation data sets $\mathcal{V}_\mathrm{unknown}$ (solid lines) with $\mathcal{V}_\mathrm{known}$ (dotted lines). The shown values are obtained by first averaging \acf*{dsc} values of all images of one pig and then averaging the mean \ac{dsc} values of the different pigs in $\mathcal{V}_\mathrm{unknown}$ and $\mathcal{V}_\mathrm{known}$. See \autoref{sec:experimental_conditions} for more details on the validation data set splits. The error curves for the superpixel and patch\_64 models are not shown for clarity and can be found in the interactive version of this plot in the supplemental material.}
\label{fig:generalization_error}
\end{figure}

\paragraph{Generalization capability} We wanted to study how well our neural network-based organ segmentation generalizes towards \varGeneralizationPhrasingS. To obtain an initial estimate for the generalization capabilities, we compared the segmentation performance achieved on the validation set containing only unseen pigs $\mathcal{V}_\mathrm{unknown}$ to those obtained on the validation set composed of unseen images of seen pigs $\mathcal{V}_\mathrm{known}$ (cf. \autoref{sec:experimental_conditions}). \autoref{fig:generalization_error_hsi} shows the average \ac{dsc} on $\mathcal{V}_\mathrm{unknown}$ and $\mathcal{V}_\mathrm{known}$, respecting the hierarchical structure of the data, for the 5 different levels of spatial granularity over all 100 epochs throughout training. Performance is generally better for $\mathcal{V}_\mathrm{known}$. For all modalities, the gap in performance between $\mathcal{V}_\mathrm{known}$ and $\mathcal{V}_\mathrm{unknown}$ is smallest for the pixel-based segmentation.

\subsection{Model input modalities}
\label{sec:results_modalities}

We wanted to investigate whether there is a benefit in using \ac{hsi} data compared to RGB and \ac{tpi} data.

\paragraph{Segmentation performance and ranking} \autoref{fig:benchmarking_box} shows that for all metrics and all models, the average segmentation performance on \ac{hsi} data is consistently better than the performance on RGB and \ac{tpi} data. While the performance gap is largest in the case of pixel-based segmentation, it decreases with an increased level of spatial granularity and is minimized in the case of image-based segmentation.  Nevertheless, a model based on \ac{hsi} data ranks better than the same model based on \ac{tpi} and RGB data regarding sampling and metric stability (\autoref{fig:ranking_dsc} and \autoref{fig:ranking_metrics}, respectively). In most cases, a model based on \ac{tpi} data ranks better than the same model based on RGB data. However, compared to the gap in performance/ranking between \ac{hsi}-based and non-\ac{hsi}-based segmentation, the difference in performance/ranking between \ac{tpi}- and RGB-based segmentation is usually smaller.

\paragraph{Generalization capability} When comparing the generalization capabilities towards \varGeneralizationPhrasingS for different input modalities in \autoref{fig:generalization_error}, it can be observed that across all modalities, performance gaps between $\mathcal{V}_\mathrm{unknown}$ and $\mathcal{V}_\mathrm{known}$ are considerably smaller for the pixel-based segmentation. Overall, the RGB pixel model yields the best generalization performance. Regarding the progress over training time, there are striking differences across modalities: while the \ac{dsc} changed relatively smoothly for \ac{tpi} and RGB data, the training was noisier for \ac{hsi} data. This holds especially true for the pixel model whereas the training was only slightly noisier for the image model. Additionally, training converged faster for the \ac{tpi} and RGB modalities whereas \ac{hsi} benefited more from longer training times.

\paragraph{Misclassifications} \autoref{fig:cm_hsi} shows the confusion matrix of the best model (image model on the \ac{hsi} modality) for the test set described in \autoref{sec:experimental_conditions}. For \varCMTotalClassesAboveThreshold out of the \varTotalClasses classes, on average more than \varCMThreshold of the organ pixels were correctly identified. For \oname{major vein}, the recall was lowest with only \varCMVenaCavaSensitivity of the organ pixels being correctly identified. Confusion matrices for the image model on the \ac{tpi} and RGB modalities are shown in \autoref{fig:cm_tpi} and \autoref{fig:cm_RGB}. To enable for a direct comparison of the organ-specific performance between the three modalities for the image model, \autoref{fig:recall} shows the recall stratified by organ and modality. Image-based segmentation of \ac{hsi} data performs better or comparable to \ac{tpi} and RGB data for all organ classes except from pancreas and major vein.

\newcommand{\useacro}[1]{%
  \ifthenelse{\equal{\detokenize{#1}}{\detokenize{RGB}}}
    {#1}
    {\acf*{#1}}%
}
\newcommand{\cmDescription}[3]{Confusion matrix of the image model on the \useacro{#1} modality for the test set. The matrix is based on the row-normalized pixel classification results from all images of one pig, i.e. the $(i,j)$-th entry depicts the percentage of pixels of the true class $i$ which were classified into the $j$-th class. This normalization was done on a porcine level yielding a confusion matrix per pig. These matrices were then averaged across pigs while ignoring non-existent entries (e.g. due to missing organs). The number in brackets indicates the standard deviation across pigs. Values $< \SI{0.1}{\percent}$ are not shown for brevity. Confusion matrices for the \useacro{#2} and \useacro{#3} modality are shown in \autoref{fig:cm_#2} and \autoref{fig:cm_#3}, respectively.}
\begin{figure*}[htb]
    \centering
    \includegraphics[width=0.93\textwidth]{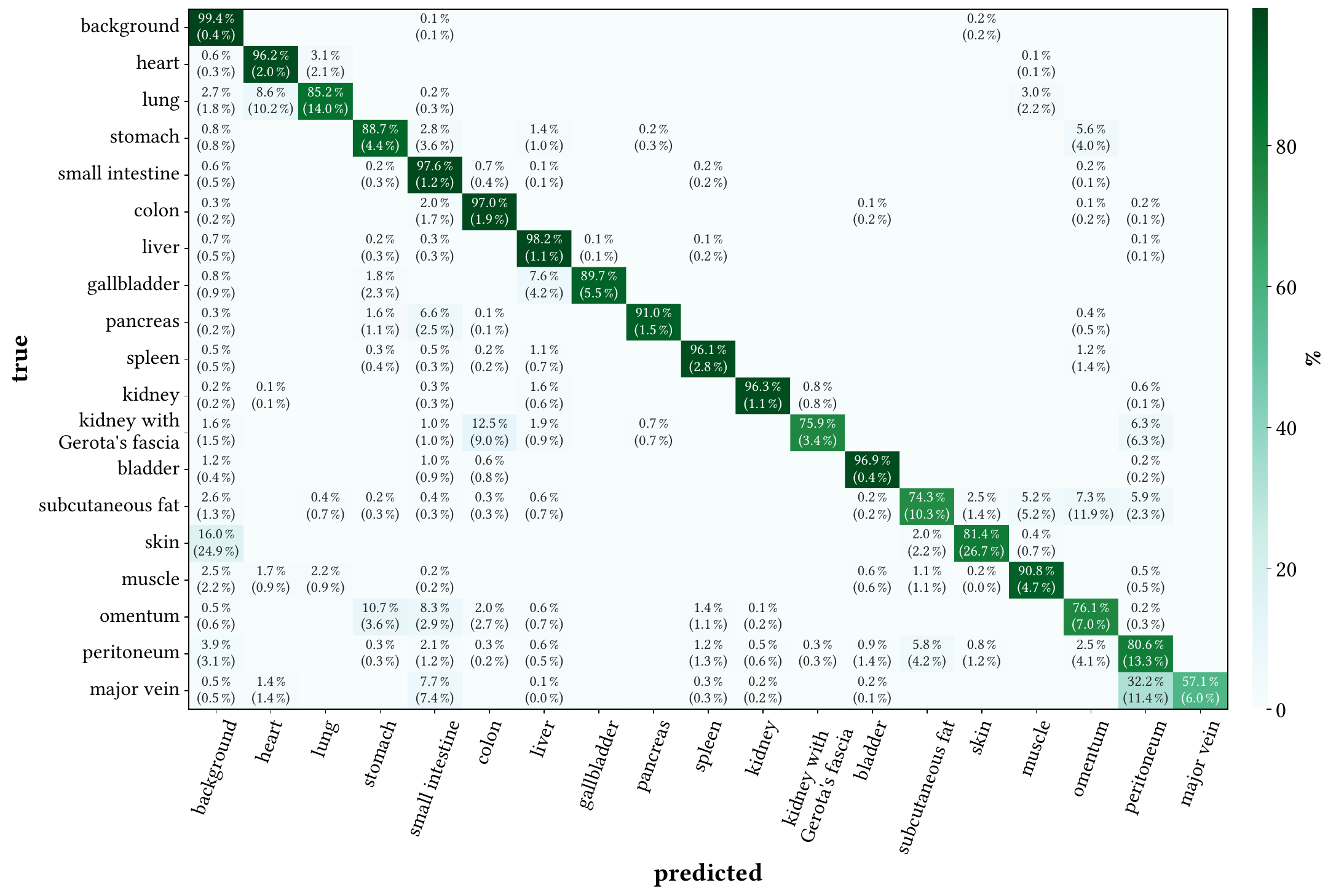}
    \caption[Confusion Matrix]{\cmDescription{hsi}{tpi}{RGB}}
    \label{fig:cm_hsi}
\end{figure*}

\section{Discussion}
\label{sec:discussion}

With the present study, we addressed two important and so far unanswered research questions in deep learning-based organ segmentation on \ac{hsi} data, namely: (1) What is the optimal spatial granularity of the input data, both with respect to segmentation performance, required amount of training data and capability to generalize towards \varGeneralizationPhrasingS? (2) Is there a benefit in using \ac{hsi} data over other modalities such as \ac{tpi} and RGB data? The key insights are:

\begin{enumerate}[noitemsep]
    \item Performance: Consistently across all our validation metrics and modalities, the segmentation performance improved with larger spatial granularity of the input data. For image-based segmentation on \ac{hsi} data, the performance of our model was comparable to that of a second medical expert.
    \item Required amount of training data: Across all studied numbers of training pigs, the image-based segmentation on \ac{hsi} data performs comparable or better than the other models.
    \item Generalization capability: Pixel-based segmentation generalizes better towards \varGeneralizationPhrasingS than image-based segmentation. For larger spatial context, the generalization capabilities are comparable across modalities.
    \item Benefit of \ac{hsi} data over \ac{tpi} and RGB data: For all models and metrics, segmentation performance on \ac{hsi} data is better than on \ac{tpi} and RGB data. However, the gap in performance decreases with increased spatial context from an improvement in the \ac{dsc} of \varPixelImprovement for the pixel, \varSuperpixelImprovement for the superpixel, \varPatchThreeTwoImprovement for the patch\_32, \varPatchSixFourImprovement for the patch\_64 to only about \varImageImprovement for the image model, compared to RGB.
\end{enumerate}

The following sections provide a detailed discussion of the results of our experiments (\autoref{sec:interpretation_results}), design choices for our models and validation (\autoref{sec:method_discussion}) as well as limitations of our study (\autoref{sec:limitations}). We further discuss the impact of our work (\autoref{sec:impact}), give an outlook on the future (\autoref{sec:outlook}) and close with our conclusions (\autoref{sec:conclusion}).

\subsection{Interpretation of results}
\label{sec:interpretation_results}

\paragraph{Segmentation performance across models}
We found that, consistently across all modalities and metrics, the segmentation performance  improved with larger spatial granularity of the input data. This raises the question whether there are practical use cases to justify using input data of small spatial context despite the apparent reduction in performance.

A potential benefit could be an improved generalization to out-of-distribution data with respect to scene geometry for a smaller input spatial granularity. We found in \autoref{fig:generalization_error} that the performance to generalize towards \varGeneralizationPhrasingS was by far best for the pixel-based models. This finding may be explained in the following way: While a pixel-based model only needs to generalize for variation in the spectral dimension, the spatial models need to additionally account for variations of the spatial context. With higher spatial context of the input data, segmentation models may thus be more sensitive to out-of-distribution scene geometries. Such changes in scene geometry may for example result from partially or completely resected organs or changes in the acquisition hardware (e.g. using a laparoscopic camera instead of an open system). Since radical changes in the scene geometry were not covered by our data set, future work should study this aspect (cf. \autoref{sec:outlook}).

Another advantage of reduced input spatial granularities is indicated by \autoref{tab:training_procedure}: Smaller input sizes generally enable larger batch sizes. Besides the advantages mentioned in \autoref{sec:training_setup} (smoother gradients and better batch statistics), this enables model improvements that benefit from a batch distribution close to the distribution of the entire data set. For example, confounders in \ac{hsi} data potentially lead to an overestimation of machine learning performance \autocite{dietrich_machine_2021}. However, some promising techniques to achieve confounder-invariant representations (e.g. metadata normalization \autocite{lu_metadata_2021}) require large batch sizes and would thus benefit from a pixel-based instead of an image-based model.

\paragraph{Ranking stability}
Across metrics, the rankings were generally in close agreement, with only one exception visible in \autoref{fig:benchmarking_box_ASD} and \autoref{fig:ranking_metrics}: Despite the smaller spatial context, superpixel-based segmentation of \ac{hsi} data ranks better than patch-based segmentation of RGB and \ac{tpi} data for the \ac{asd} metric. This may be attributed to the boundary-sensitive nature of the \ac{asd} metric. While the superpixel boundaries match the reference segmentation very well with an average lower bound for the \ac{asd} of \varSpxLimitASD if all superpixels were correctly classified (cf. \autoref{sec:limitations}), we see from the example predictions in \autoref{fig:image_examples} that sharp vertical and horizontal edges can be present in the patch-based predictions. These are due to our choice of aggregation scheme in which an image segmentation prediction is assembled from non-overlapping patches (cf. \autoref{sec:limitations}). The resulting incomplete and scattered boundaries are especially penalized by boundary-distance metrics such as the \ac{asd}, whereas well matching boundaries are rewarded (cf. \autoref{fig:ranking_metrics}).

\paragraph{Required amount of training data}
The reduced standard deviation range of the metrics with an increasing number of training pigs in \autoref{fig:dataset_size_dsc}, \autoref{fig:dataset_size_asd} and \autoref{fig:dataset_size_nsd} should be interpreted with care since pigs were always sampled without replacement and this inevitably increases the overlap of selected pigs across random selections on runs with a higher number of training pigs due to the limited number of \varTotalTrainingPigs available training pigs. For example, when randomly selecting two different sets of pigs (each of size \varMaxTrainingPigsDatasetSize) out of the \varTotalTrainingPigs training pigs without replacement, then these two sets differ only by one pig.

\paragraph{Generalization capability}
We found in \autoref{fig:generalization_error} that the organ segmentation performance is generally better for $\mathcal{V}_\mathrm{known}$, which makes sense since the gap between different images of the same pig can be expected to be smaller than the domain gap between different images of different pigs. The improved generalization capability for pixel-based models was discussed above (cf. \autoref{sec:interpretation_results}). We further saw in \autoref{fig:generalization_error} that there are striking differences in the noisiness of the training: While the \ac{dsc} is changing relatively smoothly for \ac{tpi} and RGB data, the \ac{hsi} training curve is much noisier. This holds especially true for the pixel model whereas the training was only slightly noisier for the image model. This may be attributed to the larger input feature dimension of \ac{hsi} data compared to the other modalities.

\paragraph{Segmentation performance across modalities}
We observed that increasing the input spatial context of the RGB and \ac{tpi} modalities led to the gap in segmentation performance between using \ac{hsi} data and \ac{tpi}/RGB data decreasing, potentially since the model can use additional information from the spatial context, which compensates for the lack of detail in the spectral dimension. However, the smaller performance gap for the increased spatial context may also be attributed to the quality of the provided expert annotations since the performance of our \ac{hsi} models approaches the inter-rater variability. Therefore, future work should re-evaluate the performance for different modalities on improved annotations (cf. \autoref{sec:outlook}). 

Given the possibly only small boost in segmentation performance on \ac{hsi} data compared to RGB images, further advantages and disadvantages of \ac{hsi} camera systems should be considered when choosing the optimal imaging modality for scene segmentation. Apart from the ability to differentiate tissue classes, the detailed spectral information captured by \ac{hsi} systems creates additional opportunities in surgical guidance, e.g. by revealing functional tissue information such as perfusion state or diagnosing diseased tissue \autocite{fei_chapter_2020, zhang_applications_2020}. Disadvantages of the \ac{hsi} system used in this study are the, compared to RGB imaging, long acquisition time of approximately \varAcquisitionTime, high cost and limited availability. However, \ac{hsi} is an emerging field and improvements can be expected in future realisations (e.g. video-rate intraoperative \ac{msi}, which was unavailable a decade ago, is nowadays possible \autocite{ayala_video-rate_2021}).

Regarding the comparison between \ac{tpi} and RGB data, we saw in \autoref{fig:benchmarking_box} that in most cases, a model based on \ac{tpi} data ranks better than the same model based on RGB data, indicating that the manually derived \ac{tpi} data contains relevant information for the segmentation task.

\paragraph{Misclassifications}
The confusion matrix in \autoref{fig:cm_hsi} revealed that the largest part of the confusion (\varCMVenaCavaMaxConfusion) is between \oname{peritoneum} and \oname{major vein}, which can be explained by the neighbouring relation of the two organs and the very limited amount of training data available for \oname{major vein} since it could only be imaged for \varCMVenaCavaTotalImages images (cf. \autoref{fig:dataset}) and visible parts are generally of small size with on average \varCMVenaCavaPixels. Other more frequently misclassified classes are either classes that are already difficult to annotate due to fuzzy boundaries (e.g. \oname{omentum}, \oname{peritoneum}, \oname{subcutaneous fat}) or unclear distinction (e.g. \oname{kidney with Gerota's fascia} and \oname{peritoneum}). Generally, most of the misclassifications in the confusion matrix occur between classes that are neighbours in the images (e.g. \oname{heart} instead of \oname{lung} and vice versa, \oname{stomach} instead of \oname{omentum} and vice versa, \oname{liver} instead of \oname{gallbladder}, \oname{background} instead of \oname{skin}, etc.) which may be attributed to errors in the predicted segmentation boundaries. This assumption is supported by the segmentation examples in \autoref{fig:image_examples}.

\subsection{Design choices}
\label{sec:method_discussion}

\paragraph{Class imbalances}
Our data set described in \autoref{sec:data} is highly imbalanced in terms of number of images (\autoref{fig:dataset}) as well as number of pixels per class. We therefore tested different countermeasures against imbalanced data sets during model development. More precisely, we tried weighted loss functions and oversampling strategies.

For the loss functions, we first calculated a weight $\omega^o$ for each class $o$ based on the number of pixels for this class across all images in the respective fold of the training set. We chose $\omega^o$ in such that majority classes received a low and minority classes a high weight. We then computed the loss value for the whole batch as a weighted average of the individual class losses. Even though we tried several approaches for calculating the class weights (e.g. inverse proportional weights), none of them could consistently improve the results on our validation data compared to having no weights at all.

In a similar manner, we evaluated oversampling strategies by sampling instances from minority classes more often compared to instances of majority classes based on the same weights as used for the loss function. We tried this for the pixel- and patch-based methods, but the resulting \ac{dsc} performance was always worse compared to the default sampling on the validation data.

\paragraph{Metrics}
We followed recommendations of segmentation challenges and used more than one metric to evaluate our results and to perform the ranking \autocite{ros_comparative_2021, antonelli_medical_2021}. We used an overlap-based measure (\ac{dsc}), a distance-based measure (\ac{asd}) and a measure for quantifying annotation uncertainty (\ac{nsd}). Each metric analyses specific properties of the predicted segmentation map and a model may be biased towards a metric, e.g. as discussed in \autoref{sec:interpretation_results}, the pixel model performs best under the \ac{dsc} metric (cf. \autoref{fig:ranking_metrics}), whereas the superpixel model performs best under the \ac{asd}. Hence, only a combination of multiple metrics can give insights into the overall performance of a model.

One particular design choice which affects all metrics is the case of missing classes in the prediction. Common evaluation frameworks such as MONAI usually return \texttt{nan} or \texttt{inf} values in these cases, leaving the aggregation to the user. While being less problematic for the \ac{dsc} and \ac{nsd} metric as they are bounded, the manner of handling missing classes is a crucial design choice for the \ac{asd} which is unbounded. There are several options such as ignoring these cases completely or imposing a fixed penalty which e.g. depends on the image diagonal. We set missing classes to the maximum distance of the other classes, which introduces a penalty without producing outliers. However, it has the disadvantage that the value for the missing class depends on the prediction of the other classes in the image.

Using the \ac{nsd} requires setting a (class-specific) threshold and for this, re-annotations of a subset of the images by at least one additional human rater are needed \autocite{nikolov_deep_2021}. This subset is usually small compared to the data set size (e.g. \varRaterTotalImages of \varTotalImages images in our case) since annotations of more images or even annotations from multiple annotators are often not feasible. Hence, the thresholds depend mainly on this subset and errors in these annotations have a high influence on the results. Missing classes in the re-annotation are also problematic since corresponding distances cannot be computed so that this part of an image does not contribute to the threshold. In the original formulation, this problem did not occur since an organ annotation was created separately for each known organ class \autocite{nikolov_deep_2021}. In our case, however, the annotators did not know which organ classes were present in the image.

\begin{figure}[htb]
    \centering
    \includegraphics[width=0.48\textwidth]{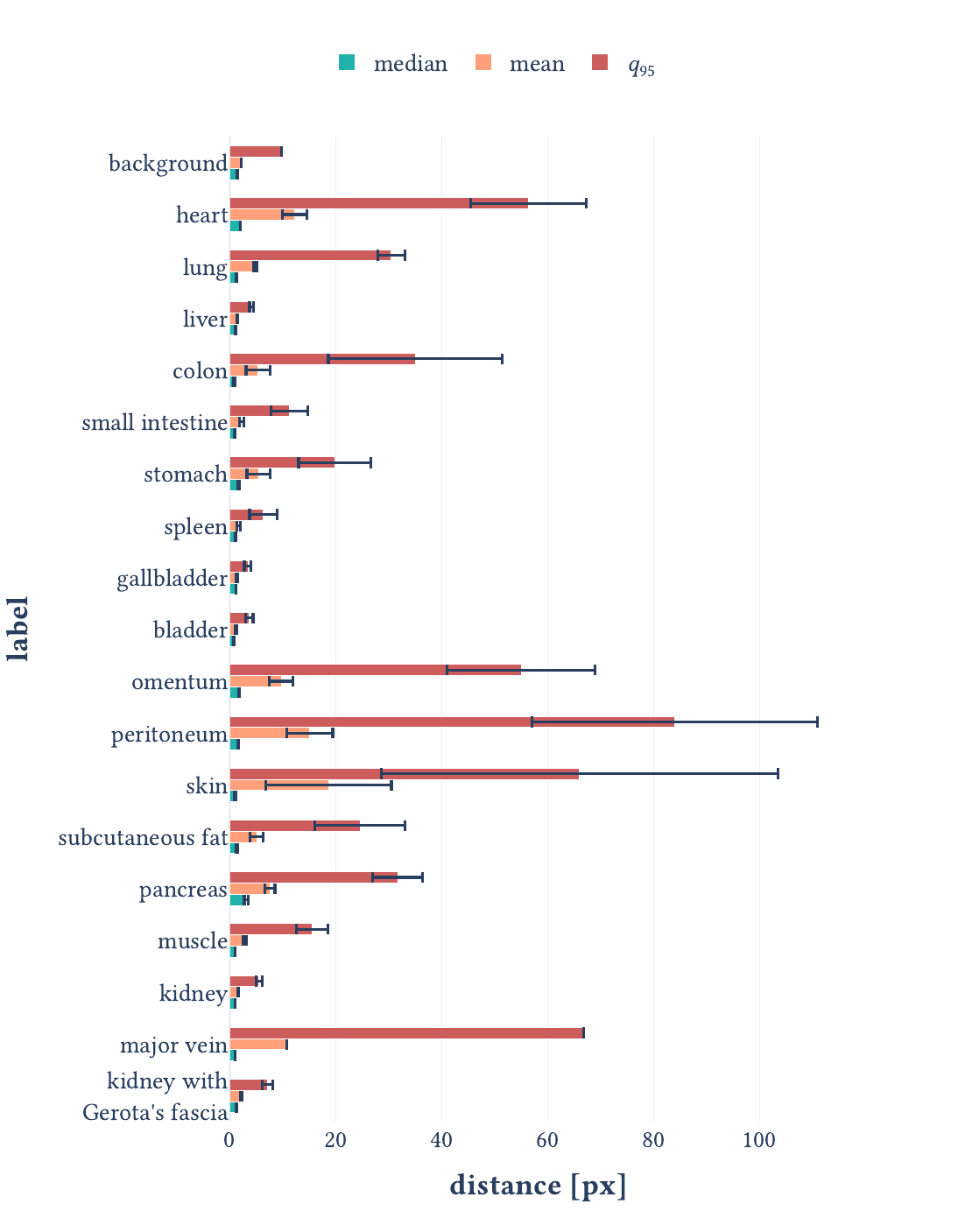}
    \caption[\ac{nsd} thresholds]{Distances which could be used as possible organ-specific thresholds $\tau^o$ for different aggregation functions (median, mean and the \SI{95}{\percent} quantile) based on the comparison of the two expert annotations. The error bars indicate 0.25 standard deviations across pigs of the aggregated values. The thresholds for the \varThresholdsAggregation aggregation correspond to the thresholds we used in our analysis (see \autoref{eq:nsd_thresholds}).}
    \label{fig:nsd_thresholds}
\end{figure}

An additional problem is the choice of the aggregation function of the class distances per organ. For each image pair which both experts annotated, we computed the distances between the two annotations for each organ, applied an aggregation function and finally averaged the aggregated values across pigs and organs (respecting the hierarchical structure). In \autoref{fig:nsd_thresholds}, we show several thresholds $\tau^o$ resulting from different aggregation functions. First, we see that there is a high variation across organ classes with e.g. large differences between the two annotations for \oname{peritoneum} and low deviations between annotations of \oname{bladder} as well as high variations across pigs (e.g. the standard deviation for the mean aggregation in case of \oname{skin} is \varThresholdsAggregationSkinFactor times higher than the mean itself). This underpins that not for every organ it is equally difficult to determine its boundary. In general, the agreement of our two expert annotations is rather low indicating that even for medical experts the decision of which pixel belongs to which organ is neither easy nor unambiguous. Second, the choice of the aggregation function is crucial for the thresholds. In their original work, Nikolov \& Blackwell \& Zverovitch et al. used the \SI{95}{\percent} quantile of the distances \autocite{nikolov_deep_2021}, but this led to very high thresholds even above \varThresholdsAggregationHigh in our case, so we decided to use the \varThresholdsAggregation instead which results in more moderate distances always below \varThresholdsAggregationLow. However, it is important to note that other aggregation methods like the median or another quantile would also have been possible.

\paragraph{Hyperparameters}
In our study, we chose default parameters whenever possible and fixed all parameters across algorithms (e.g. learning rate) or selected them based on the same criteria (e.g. memory consumption in case of the batch size). However, hyperparameters can impact the model performance and due to the different input sizes and network architectures, it is possible that our hyperparameter settings are not optimal for all algorithms. To find the optimal hyperparameter set for each algorithm, many different training runs would be required. Since training all our 15 algorithms (five models and three modalities) for all five folds already required about \varTrainingTimeTotal \ac{gpu} training time (about \SI{32}{kg} of $\mathrm{CO_2}$ equivalent if training on a GeForce GTX\texttrademark{} 2080 Ti (Nvidia Corporation, Santa Clara, USA) \autocite{lacoste_quantifying_2019}\footnote{\href{https://mlco2.github.io/impact}{https://mlco2.github.io/impact}}) just for a single hyperparameter setting, an extensive hyperparameter tuning would come at extremely high resource costs.

\begin{figure}[htb]
    \centering
    \includegraphics[width=0.49\textwidth]{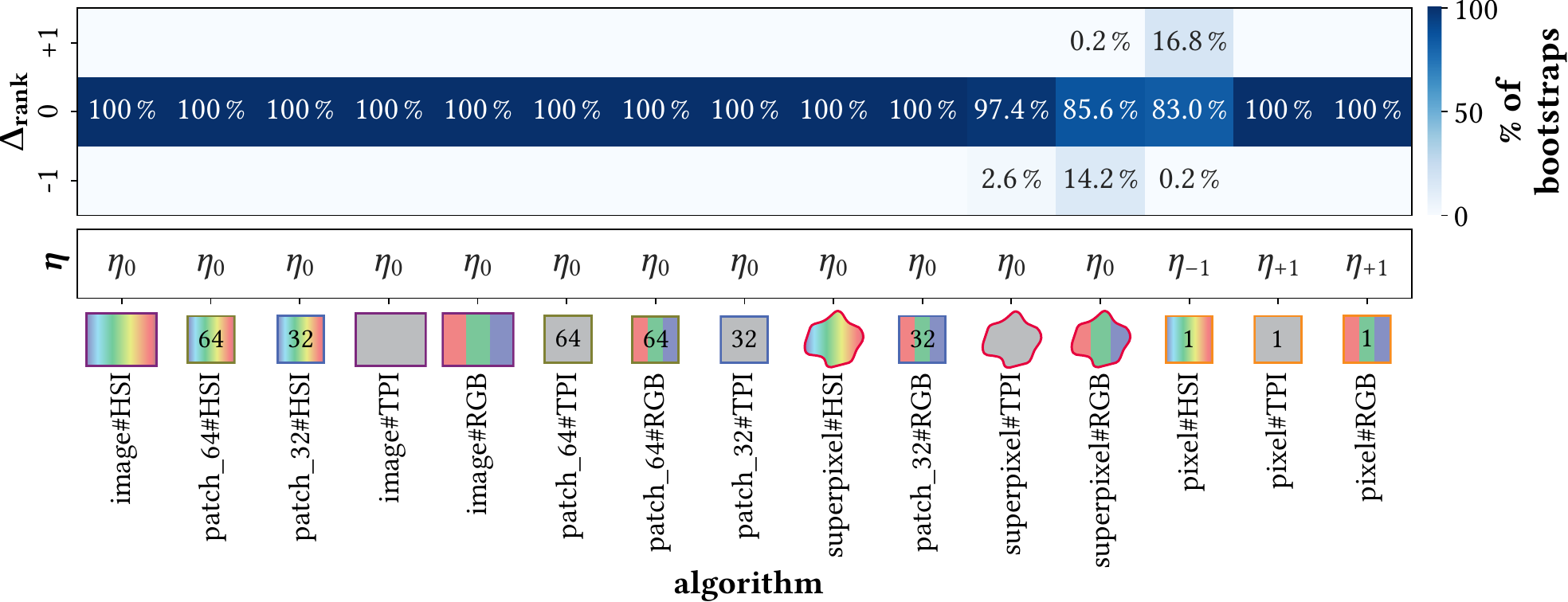}
    \caption[Differences in ranking across bootstrap samples when optimizing the learning rate]{Differences in ranking across bootstrap samples when optimizing the learning rate. For each of the 1000 bootstrap samples, the ranking based on the default learning rate $\eta_0$ is compared to the ranking based on an optimal learning rate per algorithm $\eta$ which was selected among the default ($\eta_0=\varLrDefault$), a higher ($\eta_{+1}=\varLrHigher$) and a lower ($\eta_{-1}=\varLrLower$) learning rate according to the highest average \acf*{dsc} on the validation set. The algorithms on the $x$-axis are sorted according to the overall ranking (from left to right: best to worst, cf. \autoref{fig:ranking_dsc}), which was not affected by the different choice of learning rate. The difference in the rank $\Delta_\mathrm{rank}$ was computed for each bootstrap and algorithm. Each value in the matrix depicts the fraction of bootstraps with a certain $\Delta_\mathrm{rank}$ for a certain algorithm. Fraction values of 0 are not shown to preserve clarity.}
    \label{fig:ranking_diff}
\end{figure}

To demonstrate the potential impact of our design choice on the algorithm ranking, we exemplarily evaluated a small hyperparameter search for the learning rate. We selected the learning rate since it is considered to be one of the most important hyperparameters \autocite{montavon_neural_2012}. For each algorithm, we trained two additional networks: One with a lower learning rate of $\eta_{-1}=\varLrLower$ and one with a higher learning rate of $\eta_{+1}=\varLrHigher$, compared to the default learning rate of $\eta_0=\varLrDefault$. For each algorithm individually, we determined its optimal learning rate, that is, the one learning rate among $\eta_{-1}$, $\eta_{+1}$ and $\eta_0$ that yielded the largest average \ac{dsc} on the validation data. We then repeated the overall ranking across algorithms (\autoref{fig:ranking_dsc}) on the test data, but instead of using a single fixed learning rate, the network corresponding to the optimal learning rate was selected for each algorithm. For all algorithms apart from the pixel-based models, the optimal learning rate was identical to the default learning rate. However, even for the pixel-based models, the average improvement in the \ac{dsc} was only minor (< \varLrDSCDiff for all pixel models), resulting in an overall identical to that in \autoref{fig:ranking_dsc} with only minor differences in the ranking across the different bootstrap samples as shown in \autoref{fig:ranking_diff}. This supports the validity of our study results even without an extensive algorithm-specific hyperparameter tuning.

\subsection{Limitations}
\label{sec:limitations}

\begin{figure}[ht]
\begin{subfigure}{0.48\textwidth}
  \centering
  \includegraphics[width=1\linewidth]{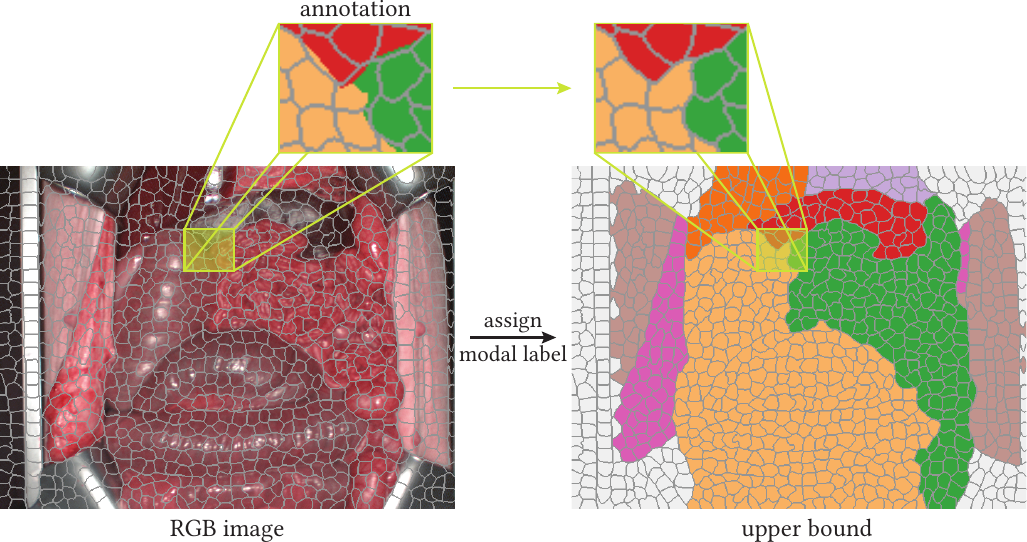}
  \caption{Sketch}
  \label{fig:spx_prediction_sketch}
\end{subfigure}
\begin{subfigure}{0.48\textwidth}
  \centering
  \includegraphics[width=1\linewidth]{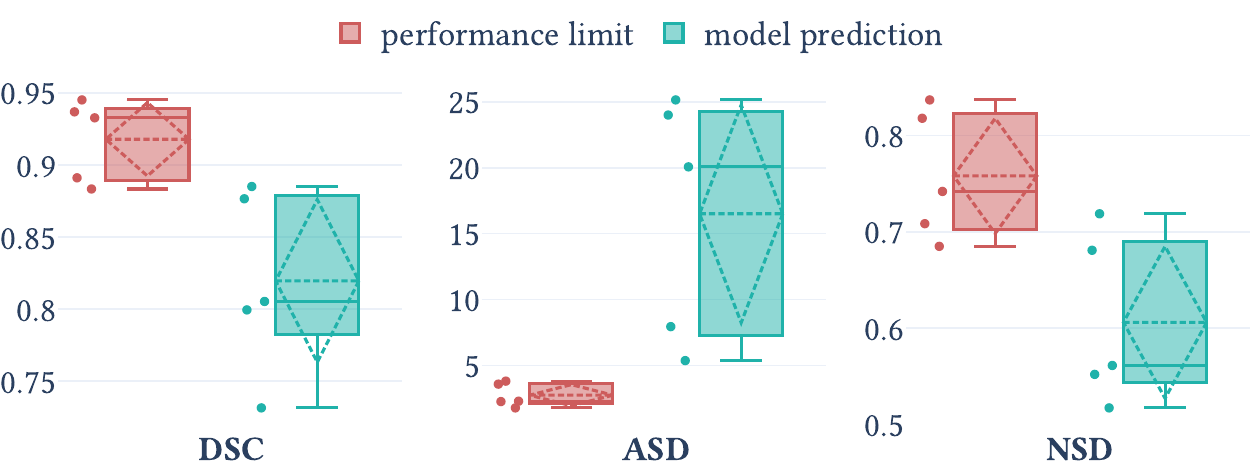}
  \caption{Evaluation}
  \label{fig:spx_prediction_results}
\end{subfigure}
\caption{Evaluation of the superpixel approach by taking into account the annotations for each superpixel which serves as performance limit. (a) sketch of the superpixel label assignment highlighting also the superpixels in the image (the same colorbar as in \autoref{fig:image_examples} applies here). (b) comparison of the prediction by either the annotation or the model prediction for three different metrics (\acf*{dsc}, \acf*{asd} and \acf*{nsd}).}
\label{fig:spx_limitations}
\end{figure}

\paragraph{Superpixels}
There are two main assumptions behind our superpixel classification approach: (1) superpixels consist of homogeneous regions with every pixel inside a superpixel belonging to the same organ class and (2) superpixel borders follow the borders of the organs (instead of crossing them). We evaluated these assumptions and created a performance limit for our superpixel model by taking the modal value of all pixel labels inside a superpixel and using this modal value as a superpixel label (\autoref{fig:spx_prediction_sketch}). This directly incorporates the annotation labels and hence serves as a performance limit for our model.

In \autoref{fig:spx_prediction_results}, we show the results of the performance limit for the different metrics on our data set. The performance limit comes closest to a perfect segmentation for the \ac{asd} with an average of \varSpxLimitASD, followed by the \ac{dsc} and \ac{nsd} with average values of \varSpxLimitDSC and \varSpxLimitNSD, respectively. The \ac{asd} is very low and has a small standard deviation because the distances between the annotation and performance limit are constrained by the superpixel size and since each superpixel contains roughly \SI{300}{\px}, large distances are unlikely (with $\sqrt{300} \approx \SI{17.32}{\px}$). For \ac{dsc} and \ac{nsd} the gap to a perfect segmentation is larger, which indicates that the superpixels are not in perfect agreement with the annotations; this can mainly be attributed to the borders between organs. It is possible that either the superpixels or the annotations (or both) are not located at the \enquote{true organ border} and any deviation leads to a reduced overlap (\ac{dsc}) or the necessity to redraw some superpixel borders to be in alignment with the annotation (\ac{nsd}). The \ac{nsd} is lower than the \ac{dsc} because the acceptable threshold $\tau^o$ is very low for some organs (cf. \autoref{fig:nsd_thresholds}) so that pixels with minor deviations already influence the \ac{nsd}.

There is a gap between the performance limit and our model predictions for all metrics with average scores of \varPerformanceSuperpixelDSC, \varPerformanceSuperpixelASD and \varPerformanceSuperpixelNSD for the \ac{dsc}, \ac{asd} and \ac{nsd}, respectively. Assuming that the features of the superpixels are discriminative enough, this indicates that our superpixel model could be improved. This is amplified by our design choices of making the models and modalities as comparable as possible without specific optimizations for a single model (e.g. when we introduced augmentations, we added them to all models). However, the image \ac{hsi} model is not too far away from the performance limit (with average values of \varPerformanceImageDSC, \varPerformanceImageASD, \varPerformanceImageNSD for the \ac{dsc}, \ac{asd} and \ac{nsd}, respectively) so that the gain in investing further in the superpixel model development is low.

\paragraph{Data set}
Non-standardized image acquisition is an important challenge in open surgery. Instead of following a standardized imaging protocol, we thus decided to acquire images in a fashion reflecting intraoperative reality (i.e. number of images and situses were decided on individually for every pig to mimic natural variations in the surgery performed). However, we are aware that this does not cover other, additional sources of variation present in a real-world open surgery setting, such as perfusion changes of certain tissue areas, presence of tissue pathologies (e.g. fatty liver), image parts being covered by body fluids (e.g. blood) and interventions being performed.

Furthermore, we found that for all studied imaging modalities, image-based models performed similarly to a second human rater. Additional performance improvements of our models may thus be limited by the quality of our reference annotations.

\paragraph{Model comparison}
We designed our models to be as comparable as possible by, for example, using the same U-Net architecture or a comparable epoch size. This was important for our study as we wanted only the size of the input and the modality to be the main sources of variation instead of model-specific design choices.

For the same reason, we also did not apply any post-processing to the network output even though models like the pixel model could benefit from this, for example via morphological operations. For inference, we restricted the spatial context of each model to its defined input size. For example, the patches which the patch\_32 model sees during inference are explicitly non-overlapping as this would increase the spatial context beyond a resolution of $32 \times 32$, making a comparison across different spatial resolutions (e.g. superpixel \textit{vs.} patch\_32) harder. However, we also see in \autoref{fig:image_examples} that this design choice produces visible artefacts, for example at patch borders in the patch-based segmentation.

Similarly, we did not introduce any \ac{hsi}-specific model variations or comparisons with existing task-specific networks to avoid biasing the results towards certain inputs. Instead, we used, for example, the same U-Net for the RGB and \ac{hsi} modalities. We see in the benchmarking of \autoref{fig:benchmarking_box} that the basic U-Net was able to leverage the additional spectral information, but the performance gap is more dominant in the patch-based and less in the image-based segmentation. This may be attributed to the U-Net leveraging the spectral information more poorly if more spatial context is available, but could also be influenced by our limitations in the annotation quality.

\subsection{Impact}
\label{sec:impact}

Many prior reports on organ segmentation in surgery have focused largely on conventional RGB imaging and minimally invasive surgery. While previous approaches to segmenting organs on \ac{hsi} data have been based on a variety of input data sizes including pixels, superpixels and patches, neither the optimal granularity of the input data has been determined nor has a clear benefit in using \ac{hsi} data over RGB data been shown. Furthermore, leveraging existing approaches to segmenting organs in an open surgery setting (in contrast to minimally invasive surgery) poses several challenges such as the large complexity and variability in the surgical scene due to non-standardized image acquisition, inter-subject variability and complex three-dimensional relationships between multiple soft tissues.

We addressed these gaps in the literature in a validation study of unprecedented size, both in number of pigs and number of classes. Despite the challenges imposed by the open surgery setting, we could segment organs with an average \ac{dsc} of \varPerformanceImageDSC in an image-based approach that is close to the inter-rater variability with an average \ac{dsc} of \varRaterInterDSC. Consistently over all input sizes, the segmentation based on \ac{hsi} outperforms the segmentation based on RGB data. We are the first to clearly show a benefit in using \ac{hsi} data compared to RGB data for organ segmentation, thus offering a foundation that further segmentation studies on \ac{hsi} data can build upon. We found that the segmentation performance decreases with less spatial context of the input data, whereas the generalization capability towards \varGeneralizationPhrasingS increases. These findings deliver valuable guidance for future work. With the inference of our image \ac{hsi} network taking only about \varInferenceTime per image on a GeForce RTX\texttrademark{} 3090 (Nvidia Corporation, Santa Clara, USA), organ segmentation maps can be obtained in real-time.

\subsection{Outlook}
\label{sec:outlook}

The following aspects should be addressed in future work:

\begin{itemize}[noitemsep]
    \item Due to our non-standardized image acquisition, some typical variations regarding situses, perspectives and number of images were covered in our data set. However, we could not cover all sources of variation that are present in real-world open surgery. Mainly, we imaged healthy and surgically unaltered pig organs and thus did not account for perfusion changes of certain tissue areas, presence of tissue pathologies (e.g. fatty liver), image parts being covered by body fluids (e.g. blood) and interventions being performed. Such out-of-distribution situations may alter the performance and resulting rankings of our algorithms. For example, changes in the scene geometry (e.g. partially or completely resected organs or changes in the acquisition hardware) may affect models with larger spatial granularity of the input data stronger than pixel-based segmentation models. These aspects should be studied in the future. Since the ultimate goal is real-world open surgery in humans, the generalization of our findings from pigs to humans should be verified.
    \item In the present study, we focused on organ segmentation. A natural next step would be to expand our work to full scene segmentation by introducing additional instruments and other non-biological object classes as well as medical classes (e.g. sutures).
    \item Our study on the intra- and inter-rater variability of our data set showed that organ segmentation is not an easy task, even for medical experts. The quality of our reference annotations may impose an upper limit to the performance of our models with our image-based models approaching the inter-rater variability. Future work may thus also focus on improving the reference annotations, e.g. by having the data annotated by multiple medical experts and seeking for consensus agreement in cases of differing annotations.
    \item We studied three different levels of spectral granularity of the input and found that the segmentation performance on \ac{hsi} data with $c=100$ channels was better than on two aggregations thereof, namely RGB images with $c=3$ and \ac{tpi} data with $c=4$ channels. A natural next step would be to study whether the segmentation performance could further be improved by a selection of $4<c<100$ input channels. Some spectral channels may be more informative than others and several studies on channel selection from \ac{hsi} data reported a performance boost in the downstream task \autocite{trajanovski_tongue_2021, ayala_band_2022}. Furthermore, knowledge of the optimal spectral channels for organ segmentation could guide the development of \ac{msi} systems that often provide higher image acquisition speed and/or spatial resolution.
\end{itemize}

\subsection{Conclusion}
\label{sec:conclusion}

In a comprehensive validation study, we showed that unprocessed \ac{hsi} data offers a huge benefit compared to RGB data or processed data of the camera provider for organ segmentation with the image-based \ac{hsi} model approaching inter-rater performance. We conclude that \ac{hsi} could become a powerful imaging modality for fully-automatic surgical scene understanding with many advantages over traditional imaging including the ability to recover additional functional tissue information.

\printacronyms

\section*{Acknowledgments and conflicts of interest}
This project has received funding from the European Research Council (ERC) under the European Union’s Horizon 2020 research and innovation programme (NEURAL SPICING, grant agreement No. 101002198) and was supported by the German Cancer Research Center (DKFZ) and the Helmholtz Association under the joint research school HIDSS4Health (Helmholtz Information and Data Science School for Health). Martin Wagner, Lena Maier-Hein and Beat P. Müller-Stich worked with the medical device manufacturer KARL STORZ SE \& Co. KG in the projects \enquote{InnOPlan} and \enquote{OP 4.1}, funded by the German Federal Ministry of Economic Affairs and Energy (grant agreement No. BMWI 01MD15002E and BMWI 01MT17001E) and \enquote{Surgomics}, funded by the German Federal Ministry of Health (grant agreement No. BMG 2520DAT82D).

The Authors declare that there is no conflict of interest.

\bibliographystyle{model2-names.bst}
\biboptions{authoryear}
\bibliography{references}

\appendix
\setcounter{figure}{0}% Start fresh with the figures counter

\section{Additional results}

\begin{figure}[h!]
\begin{subfigure}{0.48\textwidth}
  \centering
  \includegraphics[width=1\linewidth]{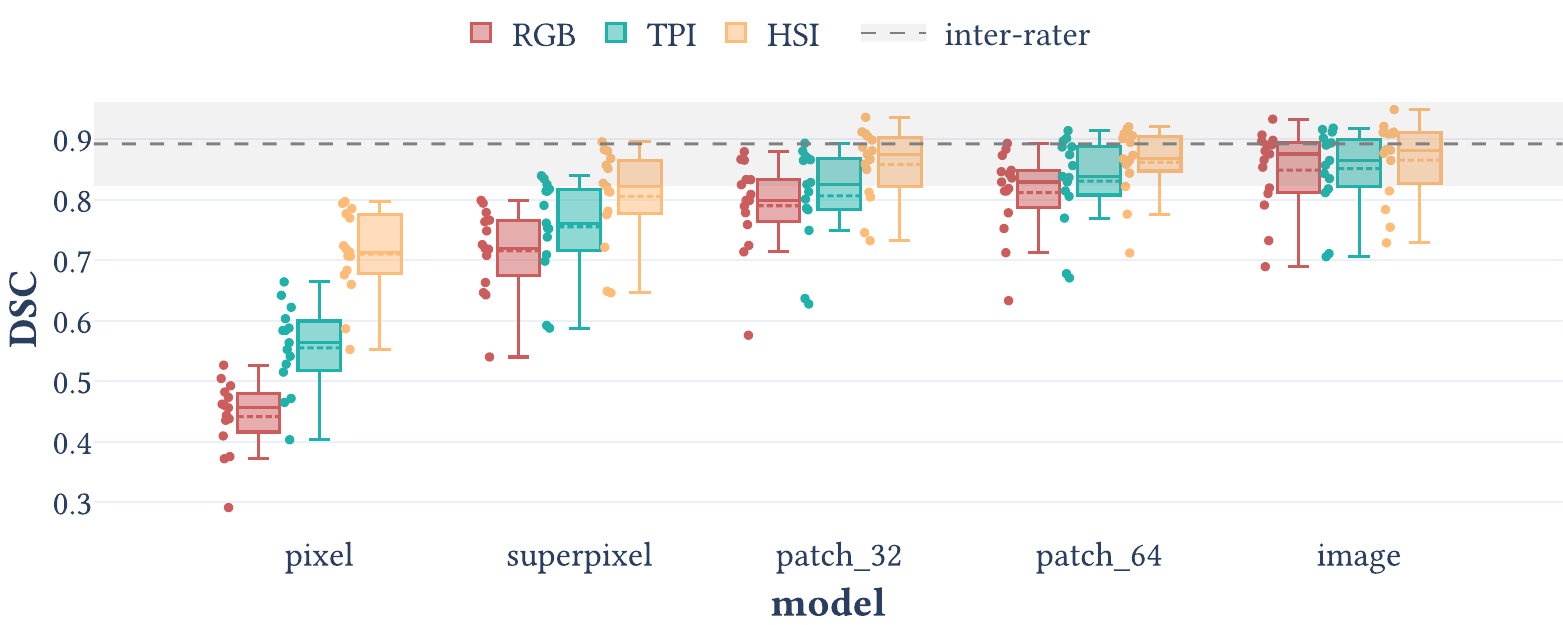}
  \caption{\Acf*{dsc}}
\end{subfigure}
\begin{subfigure}{0.48\textwidth}
  \centering
  \includegraphics[width=1\linewidth]{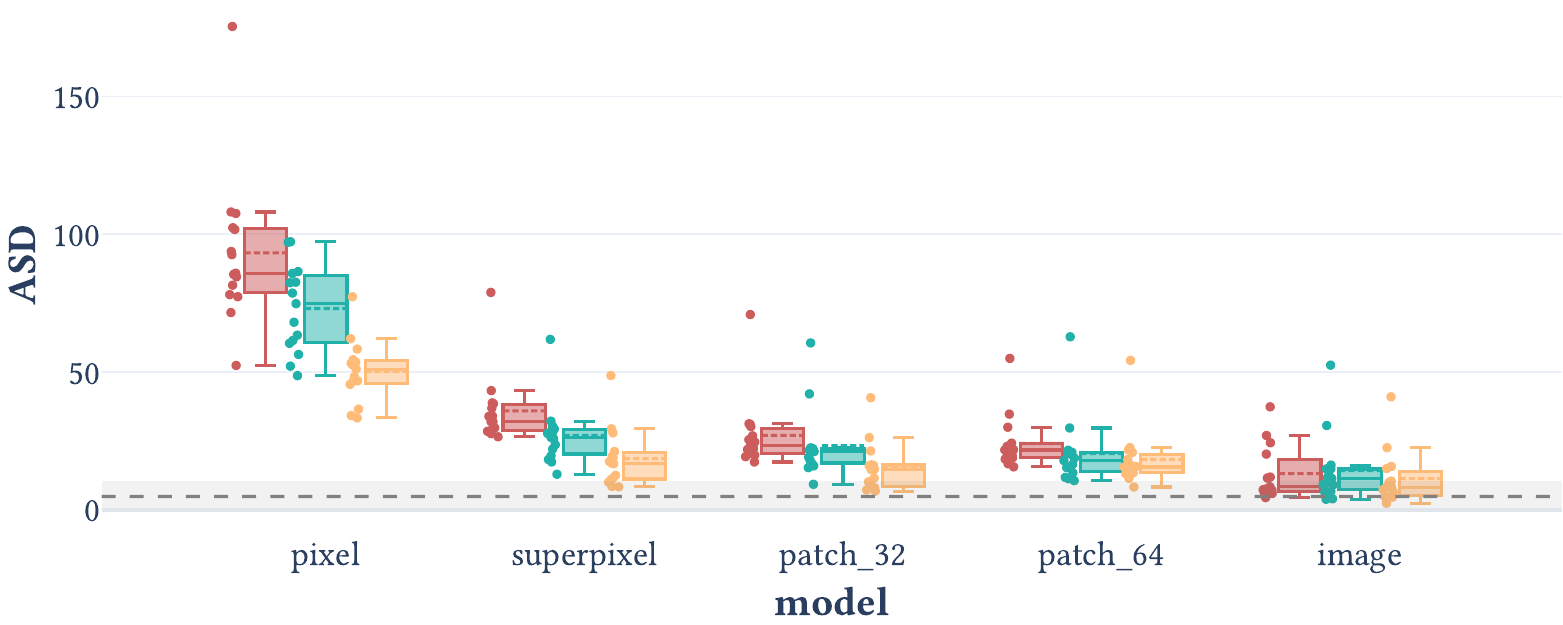}
  \caption{\Acf*{asd}}
\end{subfigure}
\begin{subfigure}{0.48\textwidth}
  \centering
  \includegraphics[width=1\linewidth]{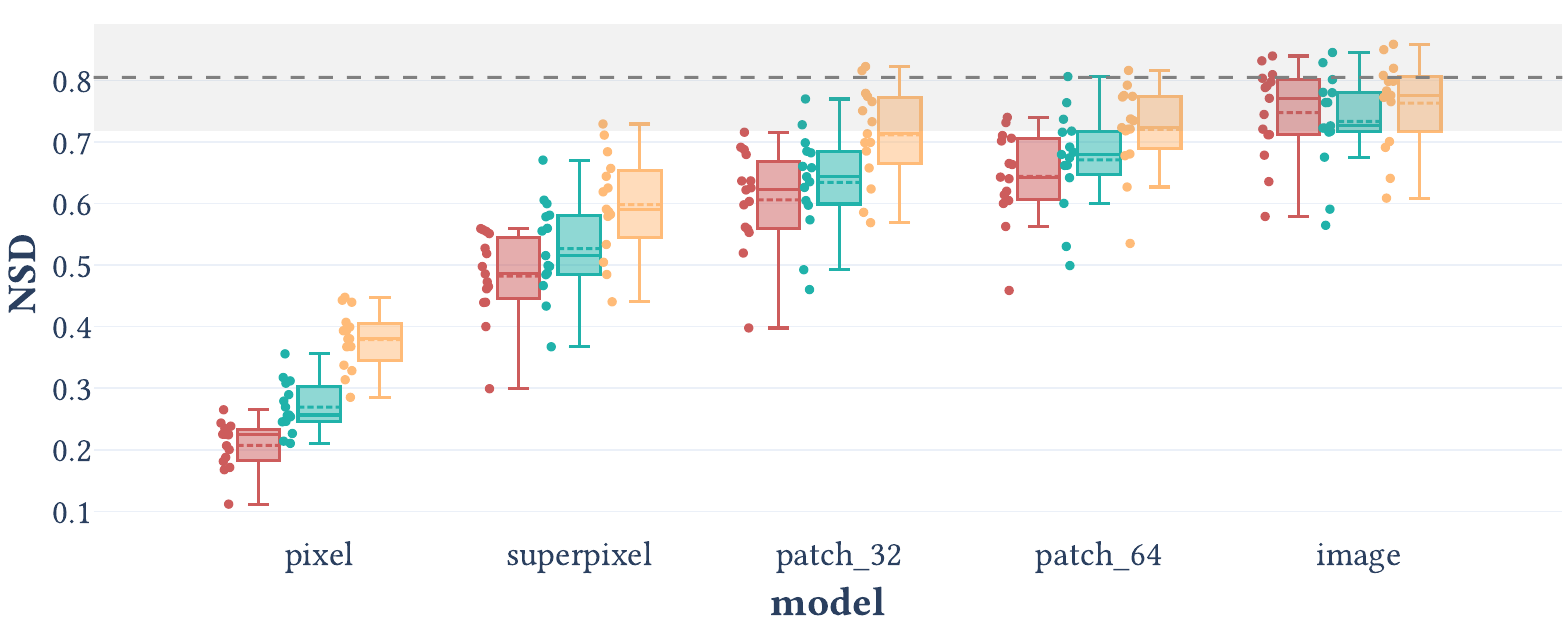}
  \caption{\Acf*{nsd}}
\end{subfigure}
\caption{\benchmarkingBoxDescription{validation} Results on the test set can be found in \autoref{fig:benchmarking_box}.}
\label{fig:benchmarking_box_val}
\end{figure}

\begin{figure}[htb]
    \centering
    \includegraphics[width=0.48\textwidth]{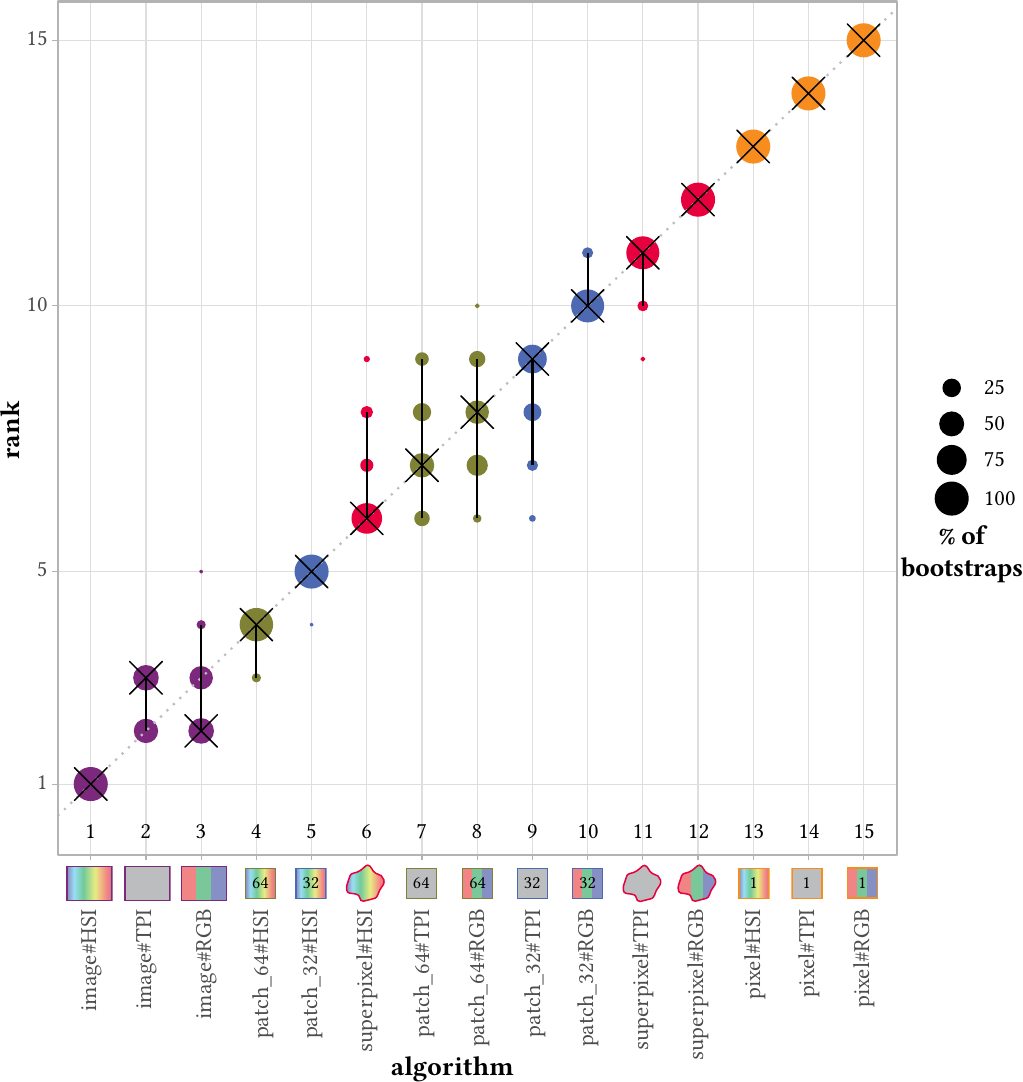}
    \caption[Ranking]{\rankingDescription{asd}{dsc}{nsd}}
    \label{fig:ranking_asd}
\end{figure}

\begin{figure}[htb]
    \centering
    \includegraphics[width=0.48\textwidth]{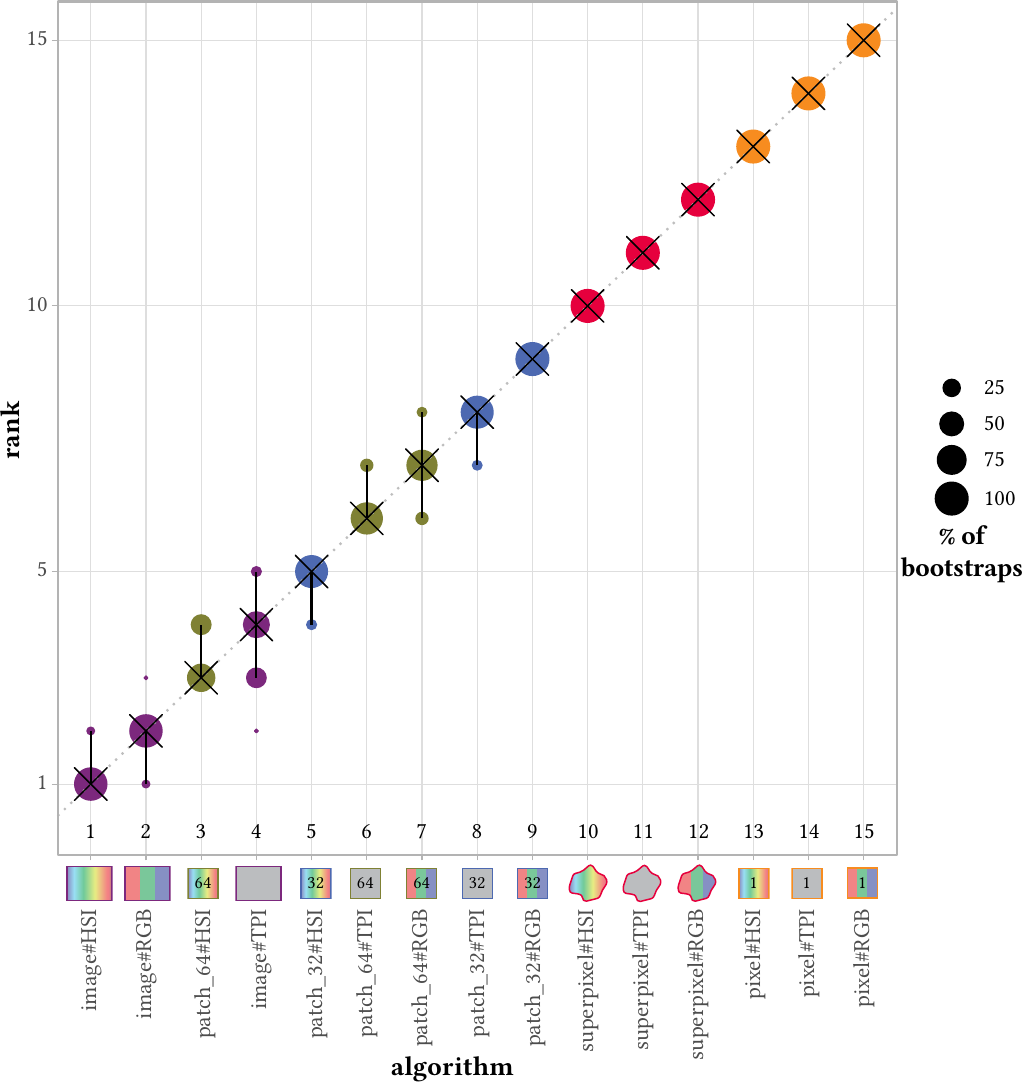}
    \caption[Ranking]{\rankingDescription{nsd}{dsc}{asd}}
    \label{fig:ranking_nsd}
\end{figure}

\begin{figure}[htb]
    \centering
    \includegraphics[width=0.48\textwidth]{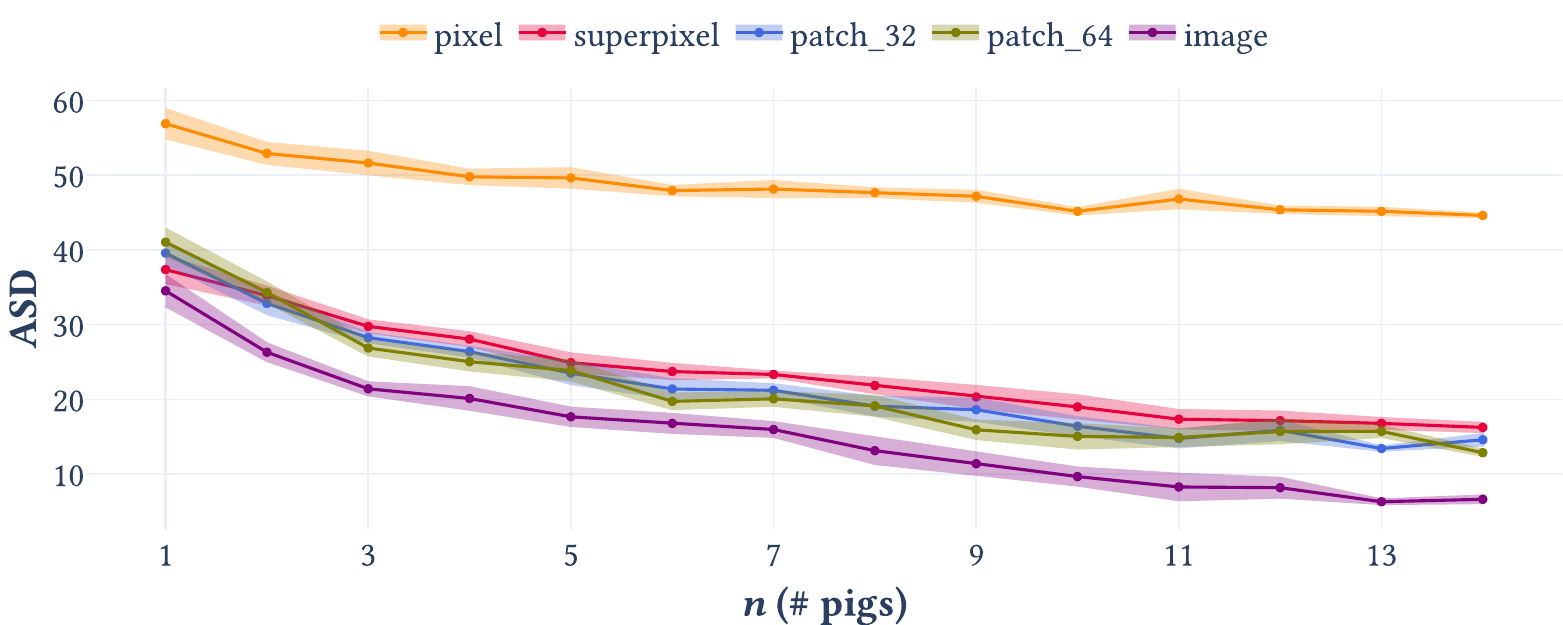}
    \caption[Data set size]{\datasetSizeDescription{asd}{dsc}{nsd}}
    \label{fig:dataset_size_asd}
\end{figure}

\begin{figure}[htb]
    \centering
    \includegraphics[width=0.48\textwidth]{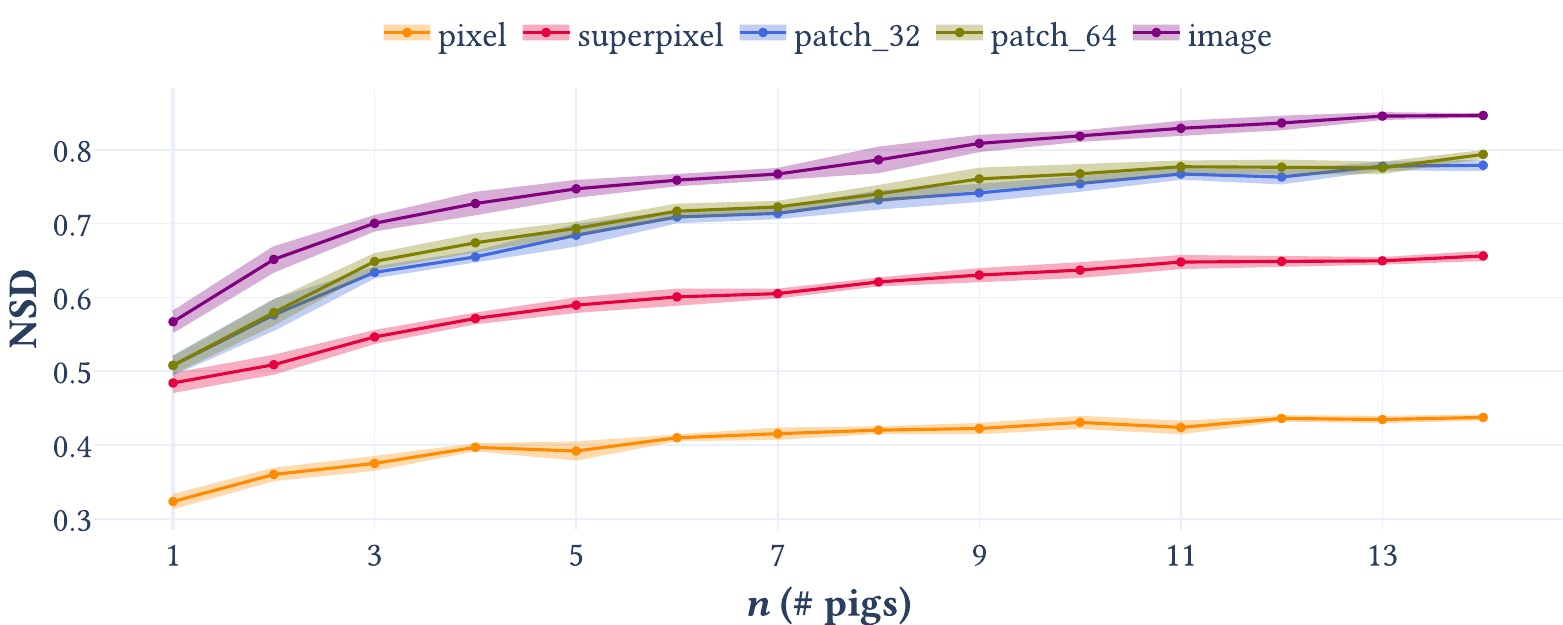}
    \caption[Data set size]{\datasetSizeDescription{nsd}{dsc}{asd}}
    \label{fig:dataset_size_nsd}
\end{figure}

\begin{figure*}[htb]
    \centering
    \includegraphics[width=0.93\textwidth]{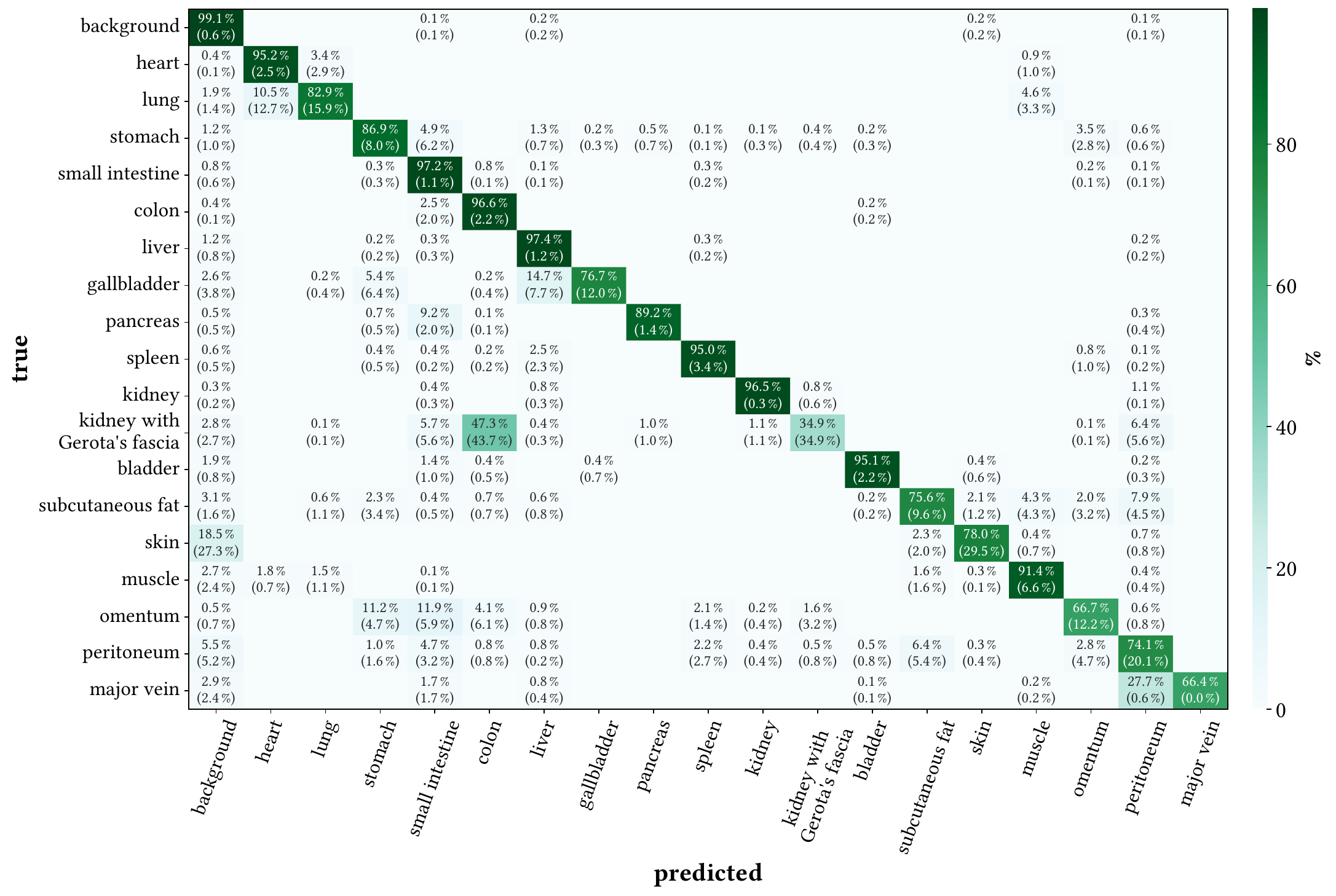}
    \caption[Confusion Matrix]{\cmDescription{tpi}{hsi}{RGB}}
    \label{fig:cm_tpi}
\end{figure*}

\begin{figure*}[htb]
    \centering
    \includegraphics[width=0.93\textwidth]{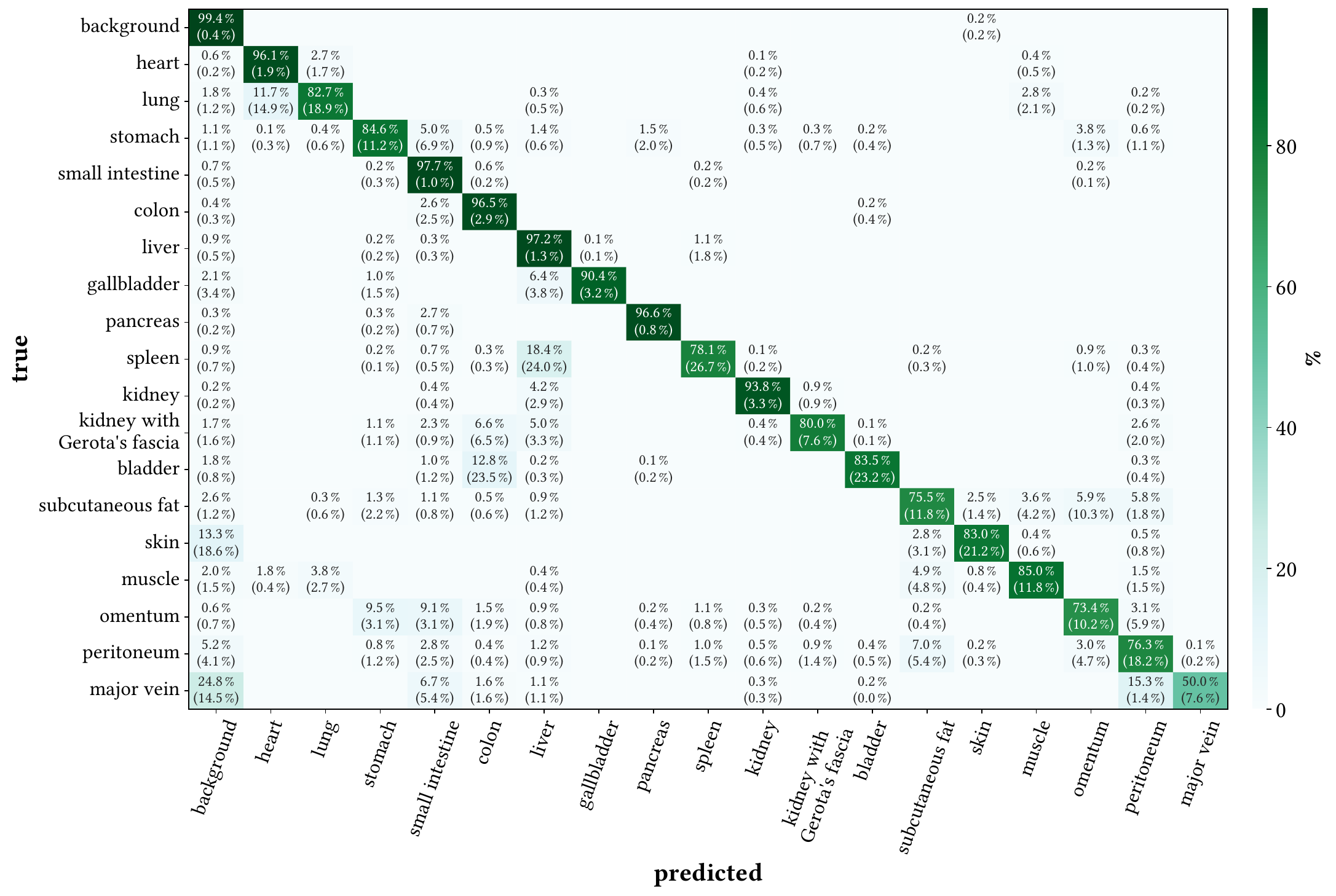}
    \caption[Confusion Matrix]{\cmDescription{RGB}{hsi}{tpi}}
    \label{fig:cm_RGB}
\end{figure*}

\begin{figure*}[htb]
    \centering
    \includegraphics[width=\textwidth]{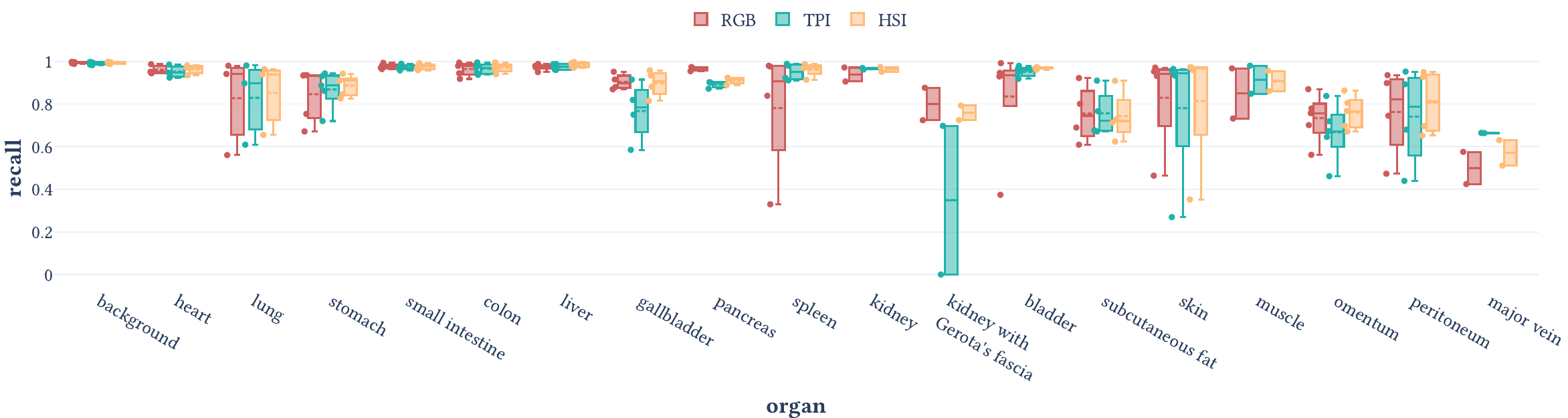}
    \caption{Recall stratified by organ and modality. The recall of the image model on the test set for three modalities (RGB, \acf*{tpi} and \acf*{hsi}) is shown, stratified by organ. Each boxplot shows the quartiles of the metric value distribution with the whiskers extending up to $1.5$ times the interquartile range, and the median and mean as solid and dashed line, respectively. Each dot represents one test pig.}
    \label{fig:recall}
\end{figure*}

\end{document}